\documentclass[a4paper,fleqn,usenatbib,useAMS]{mnras}

\usepackage{graphicx}	
\usepackage{amsmath}	
\usepackage{amssymb}	
\usepackage{multicol}   
\usepackage{bm}		
\usepackage{pdflscape}	
\usepackage{blindtext}
\usepackage{subcaption}
\usepackage{float}

\usepackage[T1]{fontenc}
\usepackage{ae,aecompl}

\usepackage{txfonts}

\title[MNRAS \LaTeX\ guide for authors]
{Slowly rotating super-compact Schwarzschild stars}

\author[C. Posada]{
Camilo Posada,$^{1}$\thanks{Contact e-mail: \href{mailto:posadaag@email.sc.edu}{posadaag@email.sc.edu}}\\
$^{1}$Department of Physics and Astronomy, University of South Carolina, 712 Main Street, Columbia SC, 29208 USA}

\date{}

\pubyear{2016}

\begin{document}
\label{firstpage}
\pagerange{\pageref{firstpage}--\pageref{lastpage}}
\maketitle

\begin{abstract}

The Schwarzschild interior solution, or `Schwarzschild star', which describes a spherically symmetric homogeneous mass with constant energy density, shows a divergence in pressure when the radius of the star reaches the Schwarzschild-Buchdahl bound. Recently Mazur and Mottola showed that this divergence is integrable through the Komar formula, inducing non-isotropic transverse stresses on a surface of some radius $R_{0}$. When this radius approaches the Schwarzschild radius $R_{s}=2M$, the interior solution becomes one of negative pressure evoking a de Sitter spacetime. This gravitational condensate star, or gravastar, is an alternative solution to the idea of a black hole as the ultimate state of gravitational collapse. Using Hartle's model to calculate equilibrium configurations of slowly rotating masses, we report results of surface and integral properties for a Schwarzschild star in the very little studied region $R_{s}<R<(9/8)R_{s}$. We found that in the gravastar limit, the angular velocity of the fluid relative to the local inertial frame tends to zero, indicating rigid rotation. Remarkably, the normalized moment of inertia $I/MR^2$ and the mass quadrupole moment $Q$ approach to the corresponding values for the Kerr metric to second order in $\Omega$. These results provide a solution to the problem of the source of a slowly rotating Kerr black hole.       

\end{abstract}




\newpage

\section{Introduction}

In classical General Relativity it is commonly accepted that the final state of complete gravitational collapse is a singular state called a `black hole' \citep{misner1973}. This object is characterized by a central space-time singularity at $r=0$ surrounded by an event horizon, a null hypersurface located at the Schwarzschild radius $R_{s}=2M$ which separates points connected to infinity by a timelike curve from those that are not. These features are a consequence of the exact solution to Einstein's field equations in the vacuum found a century ago by \cite{schwarzschild1916a} which describes the exterior space-time geometry of a spherically symmetric mass.
\\
\indent Despite the vast amount of literature [see e.g. \citep{wald2001,page2005}], the physical reality of black holes has not only generated some skepticism \citep{abramowicz2002,frolov2014,hawking2014}, but also has raised some paradoxical issues which have not been consistently solved. A pivotal one is the non-conservation of information by quantum matter falling into a black hole \citep{hawking1976breakdown}. Additionally in the original \cite{hawking1975} radiation derivation, the backward-in-time propagated mode seems to experience a large blueshift with energies larger than the Planck energy. It is expected that these highly `blue-shifted' photons would leave a non-negligible `imprint' on the spacetime geometry, making the approximation of fixed classical geometry background untenable \citep{mazur2001,mottola2011}. Moreover, the arbitrarily large values of entropy at $T_{H}\to 0$ associated with the black hole as predicted by the \cite{Bekenstein1974} formula in the classical limit $\hbar\to 0$ produces serious challenges to the foundations of quantum mechanics. It is believed that the resolution of these issues will be achieved in the framework of a consistent theory of quantum gravity. We still do not posses such a theory, therefore it is valuable to investigate alternative solutions to the aforementioned problems.
\\
\indent Alternatives have been introduced to alleviate some of the black hole paradoxes \citep{stephens1994, chapline2001, berezin2003}. In particular, we concentrate in the \emph{gravastar} (vacuum condensate gravitational star) model proposed by \cite{mazur2001, mazur22004, mazur2004}. A gravastar is basically the aftermath of the gravitational collapse of a star to the Schwarzschild radius $R_{s}$, leaving a final state characterized by a modified de Sitter interior region with negative pressure and a finite surface tension. The exterior spacetime remains the standard spherically symmetric Schwarzschild exterior solution.
\\
\indent In connection with the gravastar, \cite{mazur2015} considered the constant density Schwarzschild interior solution, or `Schwarzschild star'. It is well known that this interior solution shows a divergence in pressure when the radius of the star $R=(9/4)M$ \citep{schwarzschild1916b,buchdahl1959}. The existence of this limit in addition to the homogeneous mass approximation, considered `unrealistic', have been assumed to be sufficient reasons to exclude the Schwarzschild star from further investigation \citep{wald1984}. This complete disregard of the interior solution has left the interesting region $R_{s}<R<(9/8)R_{s}$ unexplored.
\\
\indent In a bold approach \cite{mazur2015} analyzed this `forbidden' region and found that the divergence in the central pressure is integrable through the Komar formula \citep{komar1959}, producing a $\delta$-function of transverse stresses implying a relaxation of the isotropic fluid condition on a surface of some radius $R_{0}$. In the limit when $R\to R_{s}^{+}$ from above and $R_{0}\to R_{s}^{-}$ from below, the interior region suffers a phase transition (starting at the centre) becoming one of negative pressure evoking a de Sitter spacetime. This non-singular `bubble' of dark energy which is matched to an external vacuum Schwarzschild spacetime, has zero entropy and temperature, so providing a consistent picture of a gravitational Einstein-Bose condensate, or gravastar, as the final state of complete gravitational collapse.
\\
\indent The relevance of gravastars follows from the fact that their physical properties and behaviour are governed by classical general relativity. Gravastars are being recognized as a very challenging alternative to black holes. Moreover, calculations of observational consequences of a merger of either two black holes or two gravastars in the context of gravitational waves, e.g. ringdowns \citep{chirenti2016} and afterglows \citep{abramowicz2016}, may provide methods to discriminate between black holes and gravastars.
\\
\indent Some authors \citep{lobo2006} have investigated possible sources for the interior of the gravastar, and the electrically charged case was considered by \cite{horvat2009}. The issue of stability against axial perturbations was studied by \cite{chirenti2007}. They found that gravastars are stable under axial perturbations, moreover, the quasi-normal modes of rotating gravastars deviate from those associated with a black hole. They concluded that this might help to distinguish observationally between a gravastar and a black hole. Radial and axial gravitational perturbations on thin-shell gravastars were studied by \cite{pani2009,pani2010}.
\\
\indent Perturbation theory can also be applied to the study of equilibrium configurations of slowly rotating compact objects. In a seminal paper \cite{hartle1967} provided the relativistic structure equations to determine the equilibrium configurations of slowly rotating stars to second order in the angular velocity. In Hartle's model the interior of the star is composed of a fluid characterized by a general one-parameter equation of state (EOS). This configuration is matched to a stationary and axially symmetric exterior region across a timelike hypersurface.
\\
\indent \cite{chandra1974} studied slowly rotating homogeneous masses characterized by a constant energy density, using Hartle's framework. For this configuration they solved numerically the structure equations for several values of the parameter $R/R_{s}$ where $R$ is the radius of the star and $R_{s}$ is the Schwarzschild radius. Using these solutions Chandrasekhar \& Miller calculated integral and surface equilibrium properties such as moment of inertia and mass quadrupole moment up to the Buchdahl bound. They found that the ellipticity of the star, considering constant mass and angular momentum, manifests a prominent maximum at the radius $R/R_{s}\sim 2.4$. One result of particular interest is that for a star with the `minimum possible' radius $R=(9/8)R_{s}$, the quadrupole mass moment is very close to the value associated with the Kerr metric to second order in the angular velocity.
\\
\indent Motivated by the aforementioned works, in this paper we report results of surface and integral properties of a slowly rotating Schwarzschild star in the unstudied region $R_{s}<R<(9/8)R_{s}$. These results extend those presented by \cite{chandra1974} which where considered up to the Buchdahl radius. We show that for a Schwarzschild star in the gravastar limit when $R\to R_{s}$, surface properties like moment of inertia, angular velocity and mass quadrupole moment approach the corresponding Kerr metric values. These remarkable results provide a long sought solution to the problem of the source of rotation of a slowly rotating Kerr black hole. Throughout the paper, we use geometrized units where $c=G=1$. 
\section{Schwarzschild star and gravastar limit}
\label{sect2}
The Schwarzschild interior solution corresponding to a spherical configuration with constant energy density, is discussed in standard general relativity textbooks \citep[see e.g.][]{wald1984,plebanski2006}. The starting point is a spherically symmetric spacetime in Schwarzschild coordinates
\begin{equation}\label{metric0}
ds^2 = -e^{2\nu(r)}dt^2 + e^{2\lambda(r)}dr^2 + r^2\left(d\theta^2 + \sin^2\theta d\phi^2\right).
\end{equation} 
\noindent The stress-energy tensor for a spherically symmetric fluid is given by
\begin{equation}\label{fluid}
T_{\;\;\nu}^{\mu} = 
\begin{pmatrix}
-\epsilon & 0 & 0 & 0\\
0 & p & 0 & 0\\
0 & 0 & p_{\perp} & 0\\
0 & 0 & 0 & p_{\perp}
\end{pmatrix} 
\end{equation}
\noindent where $\epsilon, p$ and $p_{\perp}$ correspond to the energy density, radial pressure and tangential pressure respectively, which are functions of $r$ only. The energy density $\epsilon$ and the pressure $p$ are related through a given one-parameter EOS. The relevant components of the Einstein equation $G^{\mu}_{\;\;\nu}=8\pi T^{\mu}_{\;\;\nu}$ are
\begin{equation}\label{einstein1} 
e^{-2\lambda}\left(2r\frac{d\lambda}{dr}-1+e^{2\lambda}\right)=8\pi\epsilon r^2,
\end{equation}
\begin{equation}\label{einstein2} 
e^{-2\lambda}\left(2r\frac{d\nu}{dr}+1-e^{2\lambda}\right)=8\pi pr^2,
\end{equation}
\noindent jointly with the energy-momentum conservation relation
\begin{equation}\label{tov} 
\nabla_{\mu}T_{\;\;r}^{\mu} = \frac{dp}{dr} + (\epsilon+p)\frac{d\nu}{dr}+\frac{2}{r}(p-p_{\perp})=0,
\end{equation}
\noindent which corresponds to the relativistic generalization of the hydrostatic equilibrium equation or Tolman-Oppenheimer-Volkoff (TOV) equation. It is conventional to introduce
\begin{equation}\label{interiorsch} 
h(r)\equiv e^{-2\lambda(r)}=1-\frac{2m(r)}{r},
\end{equation}
\noindent where the function $m(r)$ is associated with the mass within a radius $r$ and is given by the Misner-Sharp relation \citep{misner1973}
\begin{equation}\label{sharp}
m(r) = \int_{0}^{r}dr4\pi r^2\epsilon.
\end{equation}
\noindent In terms of \eqref{sharp}, \eqref{einstein2} becomes
\begin{equation}
\frac{d\nu}{dr}=\frac{m(r)+4\pi pr^3}{r\left[r-2m(r)\right]},
\end{equation}
\noindent which in the non-relativistic limit reduces to Poisson's equation $d\nu/dr=m(r)/r^2$, where $\nu(r)$ is associated to the Newtonian gravitational potential. The interior solution, or \emph{Schwarzschild star}, is matched at the boundary $r=R$ to the asymptotically flat vacuum exterior Schwarzschild solution
\begin{equation}\label{exteriorsch}
e^{2\nu(r)_{ext}} = h_{ext}(r) = 1-\frac{2M}{r},\quad\quad r\geq R
\end{equation} 
\noindent where $M$ is the total mass and $R$ is the radius of the star. The functions $p(r)$ and $m(r)$ satisfy the boundary conditions $p(R)=0$, $m(R)=M$. The interior region is modeled as an incompressible and isotropic fluid $p=p_{\perp}$ with
\begin{equation}
\epsilon = \bar{\epsilon} = \frac{3M}{4\pi R^3}=\text{const.}
\end{equation}
\noindent It is useful to define \citep{mazur2015}
\begin{equation}\label{conventions}
\epsilon\equiv\frac{3H^2}{8\pi},\quad\quad H^2=\frac{R_{s}}{R^3},  
\end{equation}
\noindent where $R_{s}=2M$ is the Schwarzschild radius. In terms of \eqref{conventions}, equations \eqref{interiorsch} and \eqref{sharp} can be solved to obtain
\begin{equation}\label{interiormh}
m(r)=\frac{4\pi}{3}\bar{\epsilon}r^3=M\left(\frac{r}{R}\right)^3,\quad h(r)=1-H^2r^2,\quad 0\leq r \leq R.
\end{equation}
\noindent From \eqref{tov} the pressure takes the form
\begin{equation}\label{interiorp}
p(r) = \bar{\epsilon}\left[\frac{\sqrt{1-H^2r^2} - \sqrt{1-H^2R^2}}{3\sqrt{1-H^2R^2}-\sqrt{1-H^2r^2}}\right].
\end{equation} 
\noindent The metric function $e^{2\nu(r)}$ for $r<R$ can be computed to give
\begin{equation}\label{interiorf}
f(r)\equiv e^{2\nu(r)} = \frac{1}{4}\left[3\sqrt{1-H^2R^2}-\sqrt{1-H^2r^2}\right]^2\geq 0.
\end{equation} 
\noindent Across the boundary of the configuration $r=R$, this function must match the exterior metric \eqref{exteriorsch}. The continuity of $f(r)$ guarantees that an observer crossing the boundary will not notice any discontinuity of time measurements. Notice that \eqref{interiorf} is regular except at some radius $R_{0}$ where the denominator in \eqref{interiorp}
\begin{equation}\label{denom}
D \equiv 3\sqrt{1-H^2R^2}-\sqrt{1-H^2r^2},
\end{equation} 
\noindent vanishes in the range $0<r<R$. Remarkably, it can be seen from \eqref{interiorp} and \eqref{interiorf} that the pressure goes to infinity at the same point where $f(r)=0$. This singular radius can be found directly from \eqref{denom} to be
\begin{equation}\label{r0}
R_{0} = 3R\sqrt{1-\frac{8}{9}\frac{R}{R_{s}}},
\end{equation} 
\noindent which is imaginary for $R/R_{s}>9/8$. In this regime, $p(r)$ and $f(r)$ are positive. Moreover, when $R\to (9/8)R_{s}^{+}$ from above, \eqref{r0} shows that $R_{0}$ approaches the real axis at $R_{0}=0$ and a divergence of the pressure appears jointly with $f(r)\to 0$. This limit value $R_{B}=(9/8)R_{s}$, or Schwarzschild-Buchdahl bound \citep{schwarzschild1916b,buchdahl1959}, fixes the maximum possible mass for a star with given radius $R$. At this radius $R_{B}$ general relativity predicts that the star cannot remain in static equilibrium. Furthermore, once the star reaches this critical point, its gravitational collapse is inevitable.\\
\begin{figure}
\centering
\includegraphics[scale=0.4]{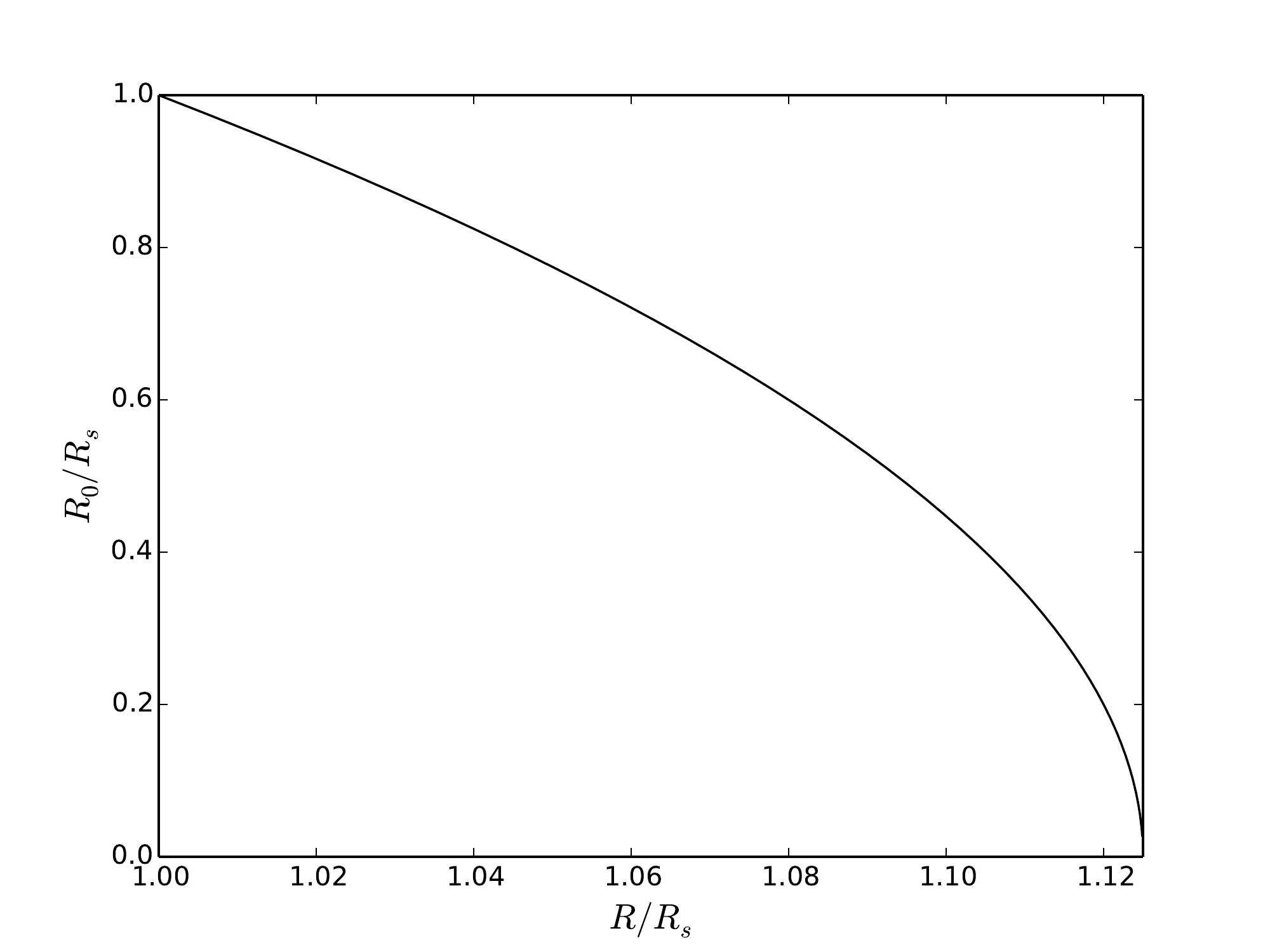}
\caption{$R_{0}$ as a function of $R$ (in units of $R_{s}$).}
\label{fig1}
\end{figure}
\begin{figure}
\centering
\includegraphics[scale=0.4]{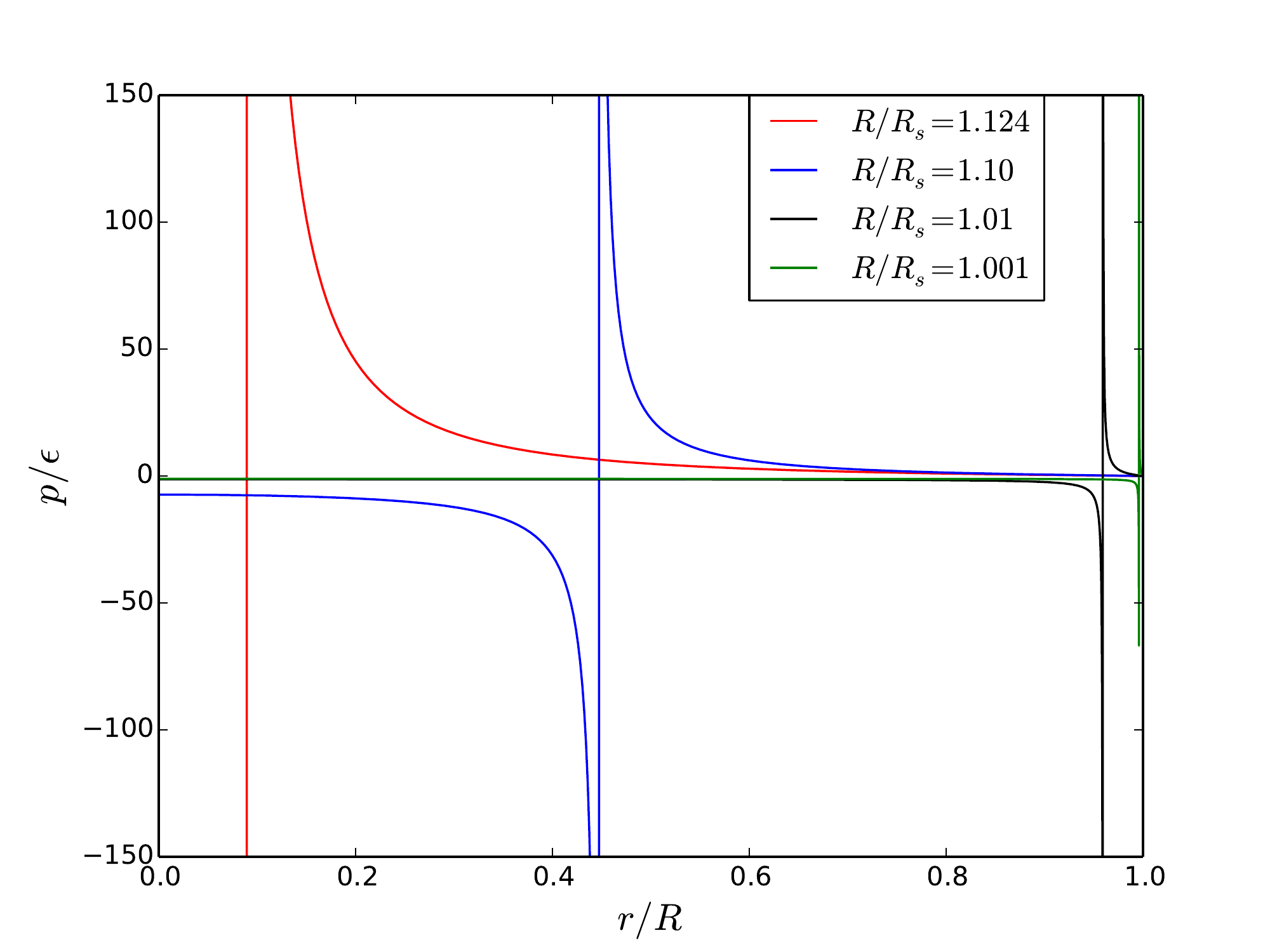}
\caption{Pressure (in units of $\epsilon$) as a function of r (in units of the stellar radius $R$) of the interior Schwarzschild solution for various values of the ratio $R/R_{s}$ below the Buchdahl bound. Notice the approach of the negative interior pressure $p\to -\epsilon$ as $R\to R_{s}^{+}$ from above and $R_{0}\to R_{s}^{+}$ from below.}
\label{fig2}
\end{figure}
\begin{figure*}
\centering
\begin{minipage}{1.0\textwidth}
\begin{subfigure}{.3\textwidth}
\includegraphics[width=\textwidth]{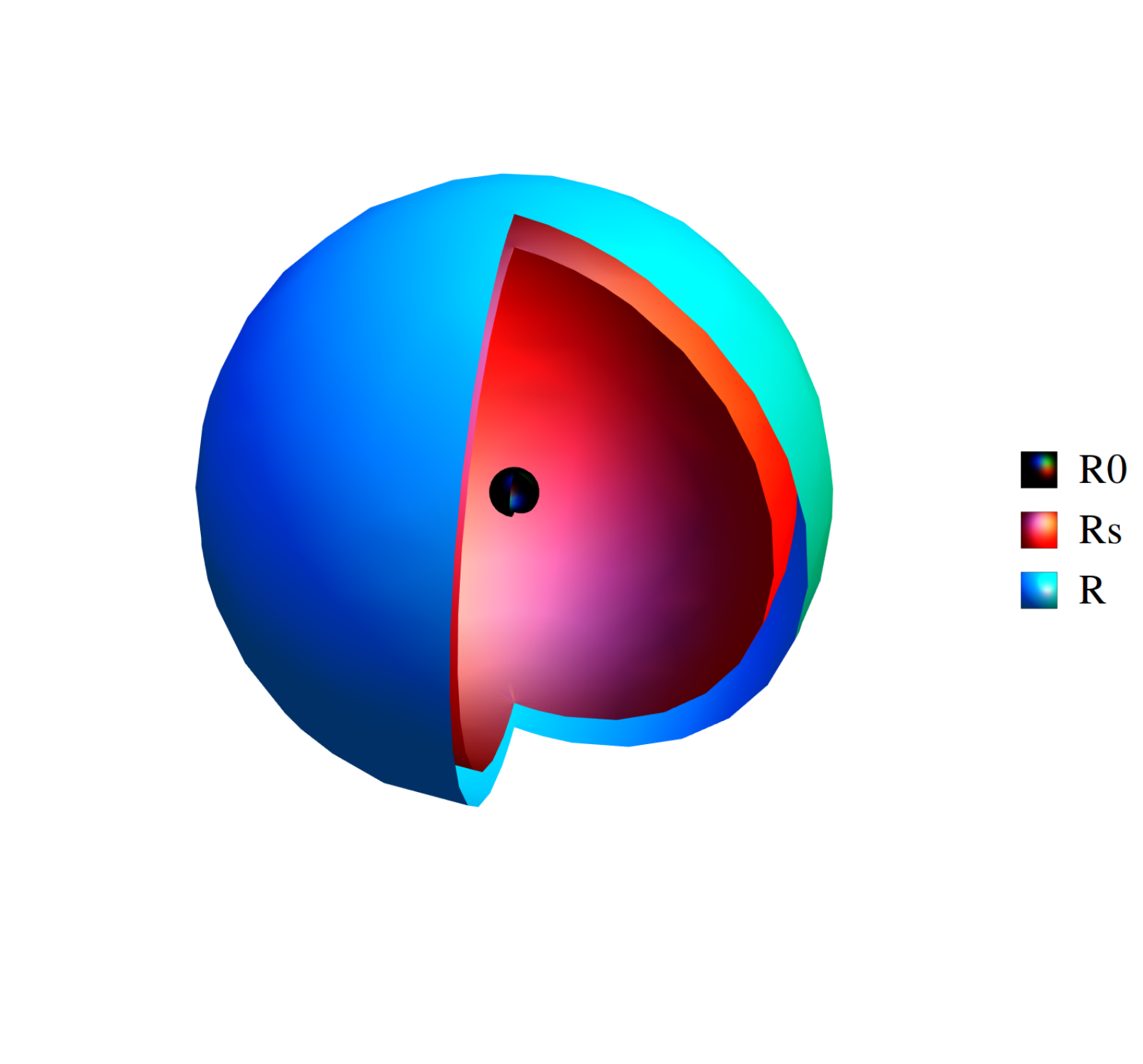}%
\caption{$R/R_{s}=1.124$}%
\label{fig3a}%
\end{subfigure}\hfill%
\begin{subfigure}{.3\textwidth}
\includegraphics[width=\textwidth]{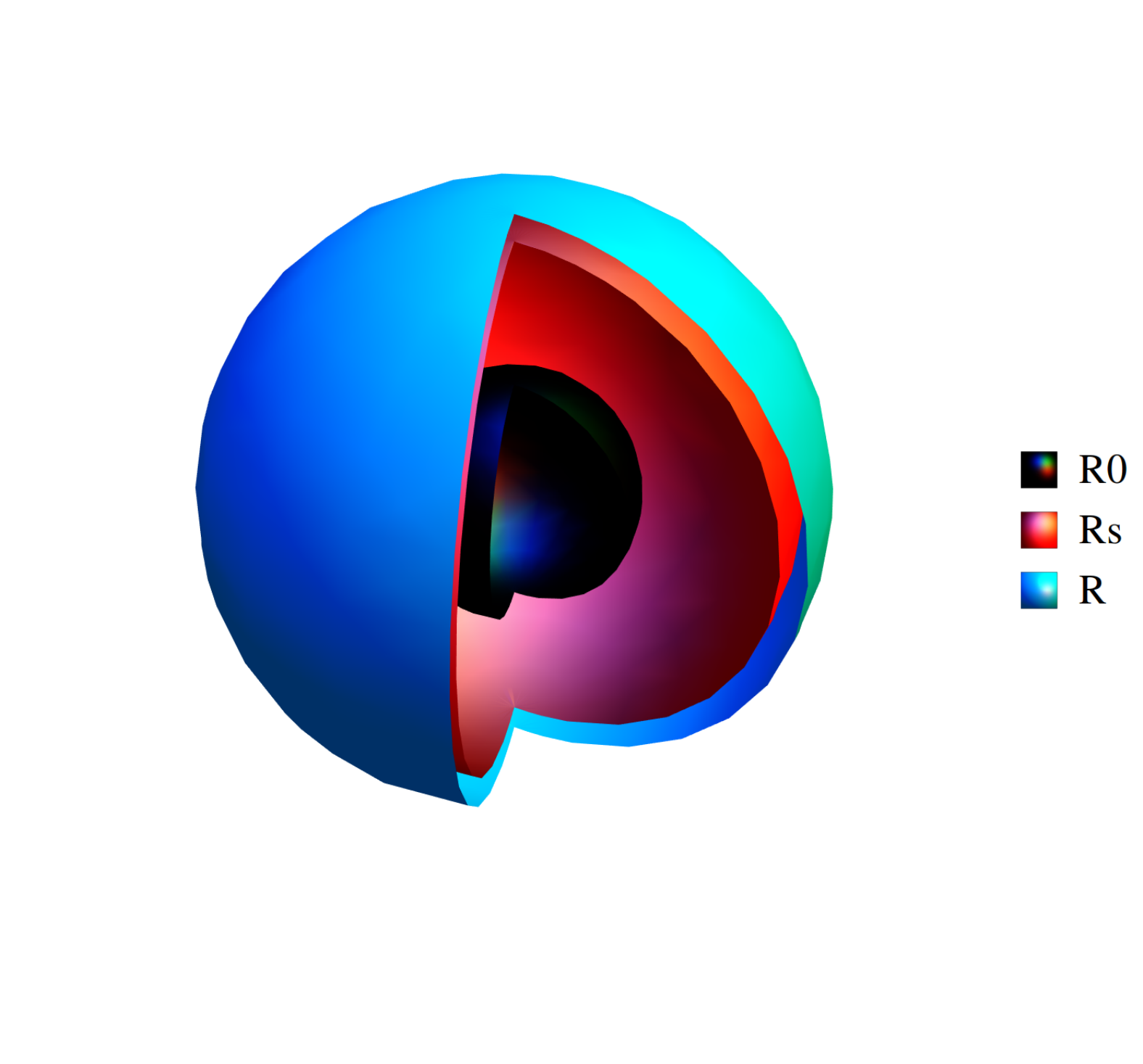}%
\caption{$R/R_{s}=1.10$}%
\label{fig3b}%
\end{subfigure}\hfill%
\begin{subfigure}{.3\textwidth}
\includegraphics[width=\textwidth]{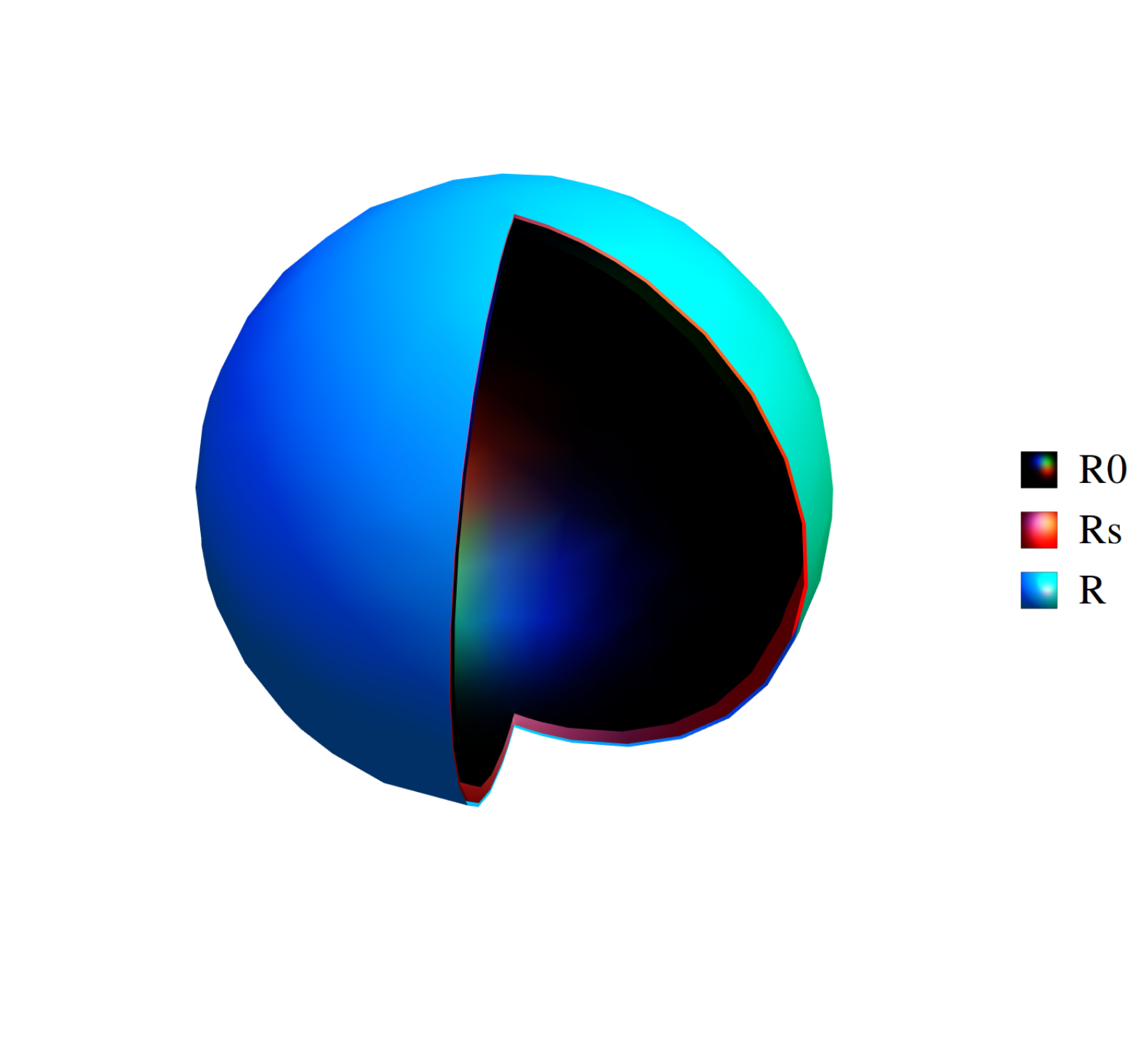}%
\caption{$R/R_{s}=1.0125$}%
\label{fig3c}%
\end{subfigure}\hfill%
\caption{Pictorial diagram of the Schwarzschild star in the regime $R_{s}<R<(9/8)R_{s}$, showing the approach of the surface of the star $R$ (Cyan) to the Schwarzschild surface $R_{s}$ (Red). The radius of the star is measured in units of the Schwarzschild radius $R_{s}$. The surface $R_{0}$ (Black) where the pressure diverges (and $f=h=0$) is shown at different stages. Figure~\ref{fig3a}, shows that $R_{0}$ emerges at the center of the star where the fluid suffers a phase transition. The region $0\leq r<R_{0}$ with negative pressure starts approaching $R_{s}$ from below, meanwhile the radius of the star $R$ approaches $R_{s}$ from above (see Fig.~\ref{fig3b}). In the gravastar limit when $R\to R_{s}^{+}$ and $R_{0}\to R_{s}^{-}$, the whole interior region is one of constant negative pressure given by a static patch of modified de Sitter spacetime with a finite surface tension (see Fig.~\ref{fig3c}). The exterior spacetime is described by the standard vacuum Schwarzschild metric. Instead of an event horizon, an infinitely thin shell forms at the Schwarzschild radius $R_{s}$ where there is a jump in pressure and the zeroes $f=h=0$ of the interior modified de Sitter and exterior Schwarzschild solutions match.}
\label{fig3}
\end{minipage}
\end{figure*}
Due to the manifestation of this divergence in pressure at the Buchdahl bound, in addition to the incompressible fluid approximation being considered artificial \citep{narlikar2010}, the Schwarzschild star solution below the Buchdahl bound has been ignored in the literature. \cite{mazur2015} analyzed the region $R_{s}<R<(9/8)R_{s}$ and they found that the zero of $D$ given by \eqref{r0} moves outwards from the origin to finite values $0<R_{0}<R$ (see Fig.\ref{fig1}). Then there emerges a region where $p(r)<0$, $f(r)>0$ and $D<0$, covering the range $0\leq r < R_{0}$. As the radius of the star keeps approaching the Schwarzschild radius from above $R\to R_{s}^{+}$, $R_{0}\to R_{s}^{-}$ from below, where $R_{0}$ is given by \eqref{r0} which corresponds to the radius of the sphere where the pressure is divergent and $f(R_{0})=0$ (see Fig.~\ref{fig3}). Analysis of \eqref{interiorp} shows that the new interior region becomes one of \emph{constant negative pressure} $p=-\epsilon$ for $r<R=R_{0}=R_{s}$ (see Fig.\ref{fig2}). In this limit, the interior metric function \eqref{interiorf} becomes
\begin{equation}\label{interior}
f(r) = \frac{1}{4}\left(1-H^2r^2\right) = \frac{1}{4}h(r)=\frac{1}{4}\left(1-\frac{r^2}{R_{s}^2}\right),\quad H=\frac{1}{R_{s}}
\end{equation} 
\noindent which is a patch of modified de Sitter spacetime. The exterior region $r>R_{s}$ remains the vacuum spherically symmetric Schwarzschild geometry \eqref{exteriorsch}, with an infinitesimal thin shell discontinuity at $R_{s}=2M$ where there is a jump in pressure and the zeroes $f=h=0$ of the interior modified de Sitter and exterior vacuum Schwarzschild spacetimes match.

Although there is no event horizon, $R=R_{s}$ is a null hypersurface. However in contrast to the black hole, the gravastar does not require the interior region $r<R_{s}$ to be trapped. Moreover, the gravastar solution with interior $p=-\epsilon$, has no entropy and zero temperature, validating its \emph{condensate state} nature.  

\cite{mazur2015} showed that the divergence in pressure at $R_{0}$ can be integrated through the Komar formula. However this integration demands that $p_{\perp}\neq p$, therefore breaking the isotropic fluid condition. Below the Buchdahl bound $R<\frac{9}{8}R_{s}$ the relation for pressure is
\begin{equation}\label{delta}  
8\pi\sqrt{\frac{f}{h}}r^2(p_{\perp}-p)=\frac{8\pi\epsilon}{3} R_{0}^3\delta(r-R_{0})
\end{equation}
\noindent indicating an anisotropy in pressure at $r=R_{0}$. It is this $\delta$-function integrable through the Komar formula, together with the relaxation of the isotropic perfect fluid condition at $r=R_{0}$ that provide a physical interpretation of the Schwarzschild star. The surface energy is found to be
\begin{equation}\label{esurf}  
E_{s}=\frac{8\pi}{3}\epsilon R_{0}^3=2M\left(\frac{R_{0}}{R}\right)^3,
\end{equation}
\noindent together with the discontinuity on the surface gravities
\begin{equation}\label{deltak}  
\delta\kappa\equiv\kappa_{+}-\kappa_{-}=\frac{R_{s}R_{0}}{R^3}
\end{equation}
\noindent provides a surface tension at $r=R_{0}$ given by
\begin{equation}\label{tension}  
\tau_{s}=\frac{MR_{0}}{4\pi R^3}=\frac{\Delta\kappa}{8\pi G}.
\end{equation}
\noindent In contrast to a black hole, \eqref{tension} corresponds to a physical surface tension (localized in an infinitesimal thin shell at $r=R_{s}$) provided by a surface energy and positive transverse pressure as determined by the Komar formula.

Notice that the Schwarzschild star solution provides an instructive limiting case of a stellar model in general relativity. Furthermore in the limit when $R\to R_{s}^{+}$ and $R_{0}\to R_{s}^{-}$, the Schwarzschild star turns out to be the non-singular gravitational condensate star or \emph{gravastar}, with a surface tension at $R_{s}$, proposed by \cite{mazur2001,mazur22004,mazur2004} as an alternative to black holes as the final state of gravitational collapse.

In the next section we will review the perturbative method developed by \cite{hartle1967}, to study equilibrium configurations of slowly rotating relativistic stars. We will apply these methods to the slowly rotating Schwarzschild star in the region $R_{s}<R<(9/8)R_{s}$.
\section{Hartle's structure equations}
\label{sect3}
In this section the equations of structure for slowly rotating masses derived by \cite{hartle1967} are summarized. The Hartle model is based on the consideration of an initially static configuration set in slow rotation. In this approximation, fractional changes in pressure, energy density and gravitational field are much less than unity. This condition implies that $R\Omega<<1$ where $R$ is the radius of the star and $\Omega$ its angular velocity. The appropriate line element for this situation is \footnote{The subscript $(0)$ in the metric functions denotes quantities in the static configuration, except for the functions $h_{0}$ and $m_{0}$ which correspond to the $l=0$ term in the harmonic expansion.}
\begin{multline}\label{axialmetric}
ds^2 = -e^{2\nu_{0}}\left[1+2h_{0}(r)+2h_{2}(r)P_{2}(\cos\theta)\right]dt^2\\
+e^{2\lambda_{0}}\left\{1+\frac{e^{2\lambda_{0}}}{r}\left[2m_{0}(r) + 2m_{2}(r)P_{2}(\cos\theta)\right]\right\}dr^2\\
+r^2\left[1+2k_{2}(r)P_{2}(\cos\theta)\right]\left\{d\theta^2 + [d\phi - \omega(r)dt]^2\sin^2\theta\right\},
\end{multline}
\noindent where $P_{2}(\cos\theta)$ is the Legendre polynomial of order 2; $(h_{0},h_{2},m_{0},m_{2},k_{2})$ are quantities of order $\Omega^2$; and $\omega$, which is proportional to the angular velocity of the star $\Omega$, is a function of $r$ that describes the dragging of the inertial frames. One can introduce a local Zero-Angular-Momentum-Observer (ZAMO), then the function $\omega(r)$ corresponds to the angular velocity of the local ZAMO relative to a distant observer. In the non-rotating case the metric \eqref{axialmetric} reduces to \eqref{metric0}.

In the coordinate system of \eqref{axialmetric}, the fluid inside the configuration rotates uniformly with four-velocity components \citep{hartle1968}
\begin{align}
u^{t} &= (-g_{tt} - 2\Omega g_{t\phi} - g_{\phi\phi}\Omega^2)^{-1/2} \notag \\
&= e^{-\nu_{0}}\left[1+\frac{1}{2}r^2\sin^2\theta(\Omega-\omega)^2e^{-\nu_{0}/2}-h_{0}-h_{2}P_{2}(\cos\theta) \right], \notag\\
u^{\phi} &= \Omega u^{t},\quad u^{r}=u^{\theta}=0.
\end{align}
\noindent It is conventional to define 
\begin{equation}
\varpi\equiv \Omega - \omega,
\end{equation}
\noindent to be the angular velocity of the fluid as measured by the local ZAMO. The magnitude of the centrifugal force is determined by this quantity which, to first order in $\Omega$, satisfies the equation
\begin{equation}\label{omegain}
\frac{d}{dr}\left(r^4j(r)\frac{d\varpi}{dr}\right)+4r^3\frac{dj}{dr}\varpi=0,
\end{equation}
\noindent with
\begin{equation}\label{j}
j(r)\equiv e^{-(\lambda_{0}+\nu_{0})}.
\end{equation}
In the exterior empty region $r>R$, $j(r)=1$ and \eqref{omegain} can be easily integrated to give
\begin{equation}\label{omegaout}
\varpi(r)=\Omega - \frac{2J}{r^3},
\end{equation}
\noindent where the constant $J$ corresponds to the angular momentum of the star \citep{hartle1967}. Equation \eqref{omegain} will be integrated outward from the origin with the boundary conditions $\varpi(0)=\varpi_{c}=const.$, and $d\varpi/dr=0$. The value of $\varpi_{c}$ is chosen arbitrarily. Once the solution on the surface is found, one can determine the angular momentum $J$ and the angular velocity $\Omega$ through the formulas
\begin{equation}\label{momentumatR}
J=\frac{1}{6}R^4\left(\frac{d\varpi}{dr}\right)_{r=R},\quad \Omega = \varpi(R) + \frac{2J}{R^3}.
\end{equation}
The angular momentum is related linearly to $\Omega$ through the relation $J = I\Omega$, where $I$ is the relativistic moment of inertia.

Additionally, due to the rotation, the star will deform carrying with it changes in pressure and energy density given by \citep{hartle1968}
\begin{align}
p+(\epsilon + p)\left[\delta p_{0}+\delta p_{2} P_{2}(\cos\theta)\right]\equiv p+\Delta P\\
\epsilon+(\epsilon + p)(d\epsilon/dp)\left[\delta p_{0}+\delta p_{2} P_{2}(\cos\theta)\right]\equiv \epsilon+\Delta\epsilon 
\end{align} 
\noindent where $\delta p_{0}$ and $\delta p_{2}$ are functions of $r$, proportional to $\Omega^2$, which correspond to perturbations in pressure and energy density. 
The spherical deformations can be studied from the $l=0$ equations with the condition that the central energy density is the same as in the static configuration. The relevant expressions are
\begin{equation}\label{m0int}
\frac{dm_{0}}{dr}=4\pi r^2(\epsilon+p)\frac{d\epsilon}{dp}\delta p_{0} + \frac{1}{12}r^4j^2\left(\frac{d\varpi}{dr}\right)^2-\frac{1}{3}r^3\varpi^2\frac{dj^2}{dr},
\end{equation}
\begin{equation}\label{h0int}
\frac{dh_{0}}{dr}=-\frac{d}{dr}\delta p_{0} + \frac{1}{3}\frac{d}{dr}\left(r^2e^{-2\nu_{0}}\varpi^2\right).
\end{equation}
These equations will be integrated outward from the origin, where the boundary conditions $h_{0}(0)=m_{0}(0)=0$ must be satisfied. In this approximation, the slowly rotating configuration will have the same central pressure as in the static case. In the exterior region, \eqref{m0int} and \eqref{h0int} can be integrated explicitly to give
\begin{equation}\label{m0out}
m_{0}=\delta M - \frac{J^2}{r^3},
\end{equation}
\begin{equation}\label{h0out}
h_{0}=-\frac{\delta M}{r-2M_{0}} + \frac{J^2}{r^3(r-2M_{0})},
\end{equation}
\noindent where $M_{0}$ corresponds to the total mass of the star and $\delta M$ is an integration constant which is associated to the change in mass due to the rotation. This constant $\delta M$ can be found by matching the interior and exterior solutions for $h_{0}$ at the boundary $r=R$.

Recently \cite{reina2015} revisited Hartle's framework within the context of the modern theory of perturbed matchings. They found that the perturbative functions at first and second order are continuous across the boundary of the configuration except when the energy density is discontinuous there. In this particular case, the discontinuity in the radial function $m_{0}$ at the boundary is proportional to the energy density there. Furthermore, Reina and Vera showed that the manifestation of this jump in the perturbative function $m_{0}$ induces a modification to the original change of mass \eqref{m0out}, which is given by \citep{reina2016}

\begin{align}\label{deltamc}
\delta M &= \delta M^{H} + \delta M^{C} \notag\\
&= \left[m_{0}(R)+\frac{J^2}{R^3}\right]+4\pi\frac{R^3}{M_{0}}(R-2M_{0})\epsilon(R)\delta p_{0}(R).
\end{align}

\noindent where $\delta M^{H}$ corresponds to the original change of mass \eqref{m0out} and $\delta M^{C}$ is the correction term. \cite{reina2016} points out that this correction is relevant in configurations where the energy density does not vanish at the boundary, for instance for homogeneous masses. We shall consider the corrected expression \eqref{deltamc} in our computations.

The quadrupole deformations of the star are computed from the integrals of the $l=2$ equations which give
\begin{multline}\label{v2in}
\frac{dv_{2}}{dr}= -2\frac{d\nu_{0}}{dr}h_{2} +\\
\left(\frac{1}{r}+\frac{d\nu_{0}}{dr}\right)\left[\frac{1}{6}r^4j^2\left(\frac{d\varpi}{dr}\right)^2-\frac{1}{3}r^3\varpi^2\frac{dj^2}{dr}\right],
\end{multline}
\begin{multline}\label{h2in}
\frac{dh_{2}}{dr}= -\frac{2v_{2}}{r\left[r-2m(r)\right](d\nu_{0}/dr)} \\
+ \left\{-2\frac{d\nu_{0}}{dr}+\frac{r}{2\left[r-2m(r)\right](d\nu_{0}/dr)}\left[8\pi(\epsilon+p)-\frac{4m(r)}{r^3}\right]\right\}h_{2}\\
+\frac{1}{6}\left[r\frac{d\nu_{0}}{dr}-\frac{1}{2\left[r-2m(r)\right](d\nu_{0}/dr)}\right]r^3j^2\left(\frac{d\varpi}{dr}\right)^2\\
-\frac{1}{3}\left[r\frac{d\nu_{0}}{dr}+\frac{1}{2\left[r-2m(r)\right](d\nu_{0}/dr)}\right]r^2\varpi^2\frac{dj^2}{dr},
\end{multline}
\begin{equation}\label{int6}
m_{2}=\left[r-2m(r)\right]\left\{-h_{2}-\frac{1}{3}r^3\left(\frac{dj^2}{dr}\right)\varpi^2+\frac{1}{6}r^4j^2\left(\frac{d\varpi}{dr}\right)^2\right\}
\end{equation}
\noindent where $v_{2}=h_{2}+k_{2}$. These equations will be integrated outward from the center, where $h_{2}=v_{2}=0$. Outside the star, \eqref{v2in} and \eqref{h2in} are integrated analytically 
\begin{equation}\label{h2out}
h_{2}(r)=J^2\left(\frac{1}{M_{0}r^3}+\frac{1}{r^4}\right)+KQ_{2}^{\;2}\left(\frac{r}{M_{0}}-1\right),
\end{equation}
\begin{equation}\label{v2out}
v_{2}(r)=-\frac{J^2}{r^4}+K\frac{2M_{0}}{\left[r(r-2M_{0})\right]^{1/2}}Q_{2}^{\;1}\left(\frac{r}{M_{0}}-1\right),
\end{equation}
\noindent where $K$ is an integration constant which can be found from the continuity of the functions $h_{2},v_{2}$ at the boundary, and $Q_{n}^{\;m}$ are the associated Legendre functions of the second kind. The constant $K$ in \eqref{h2out} and \eqref{v2out} is related to the mass quadrupole moment of the star, as measured at infinity, through the relation \citep{hartle1968}
\begin{equation}\label{quadrupole}
Q = \frac{J^2}{M_{0}} + \frac{8}{5}KM_{0}^3.
\end{equation}
\noindent Due to the rotation, the surface of the Schwarzschild star will be deformed from the spherical shape it has in the static case, preserving the same central density. The modified radius of the slowly rotating isobaric surface is given by
\begin{equation}\label{newr}
r(\theta)=r_{0}+\xi_{0}(r_{0})+\xi_{2}(r_{0})P_{2}(\cos\theta),
\end{equation}
\noindent where $r_{0}$ corresponds to the radius of the spherical surface in the non-rotating case, and the deformations $\xi_{0}$ and $\xi_{2}$ satisfy
\begin{equation}\label{deform}
\delta p_{0}=-\left(\frac{1}{\epsilon+p}\frac{dp}{dr}\right)_{0}\xi_{0}(r_{0}),\quad \delta p_{2}=-\left(\frac{1}{\epsilon+p}\frac{dp}{dr}\right)_{0}\xi_{2}(r_{0}).
\end{equation}
\noindent The ellipticity of the isobaric surfaces can be computed from \citep{thorne1971,miller1977}
\begin{equation}\label{ellip}
\varepsilon(r)=-\frac{3}{2r}\left[\xi_{2}(r)+r(v_{2}-h_{2})\right],
\end{equation}  
\noindent which is correct to order $\Omega^2$. 
\section{Structure equations for the Schwarzschild star}
\label{sect4}
In a seminal paper \cite{chandra1974} studied slowly rotating homogeneous masses using Hartle's framework. In that paper, the structure equations were integrated numerically and surface properties were computed for several values of the parameter $R/R_{s}$, where $R$ is the radius of the star and $R_{s}$ is the Schwarzschild radius. This procedure represented a quasi-stationary contraction of the star \citep{miller1977}. We will refer to the relevant equations of that paper prefixed by the letters CM, for example (CM.1) indicates equation (1) of \citep{chandra1974}.\\

The geometry of the Schwarzschild star was discussed in \ref{sect2}. In order to facilitate the numerical integrations, it is useful to introduce the coordinates (CM.35)
\begin{equation}\label{transf}
r=(1-y^2)^{1/2},\quad y_{1}^2=1-\frac{R^2}{\alpha^2}=1-H^2R^2.
\end{equation}
\noindent Here $r$ is being measured in the unit $\alpha=1/H$, where $H$ is given by \eqref{conventions}. In terms of these variables the Schwarzschild star solution \eqref{interiorsch}, \eqref{interiormh}, \eqref{interiorp} and \eqref{interiorf} takes the form
\begin{equation}\label{interior1}
e^{\lambda_{0}}=\frac{1}{y},\quad e^{\nu_{0}}=\frac{1}{2}\lvert 3y_{1}-y\rvert,\quad p=\frac{y-y_{1}}{3y_{1}-y}\epsilon,
\end{equation}
\begin{equation}\label{interior2}
j=\frac{2y}{\lvert{3y_{1}-y\rvert}},\quad \frac{2m(r)}{r}=1-y^2.
\end{equation}
\noindent Notice that, in contrast to (CM.34), here we consider the modulus of $(3y_{1}-y)$ in \eqref{interior1} in harmony with the fact that the metric element $e^{2\nu_{0}}$ in \eqref{interiorf} is a perfect square making it always a positive quantity. Notice that the function $e^{\nu_{0}}$ is negative when $3y_{1}<y$ which occurs in the region below the Buchdahl bound. Therefore, in order to investigate the region $R_{s}<R<(9/8)R_{s}$ it is important to specify the modulus condition. This specification was taken into account in the code to compute the numerical solutions. To facilitate computations, we define the quantity
\begin{equation}\label{modulok}
k=\lvert 3y_{1}-1\rvert,
\end{equation}
which is always positive as it is required for the analysis of the region $R_{s}<R<(9/8)R_{s}$. It is also advantageous to introduce the coordinate (CM.39)
\begin{equation}\label{newx}
x\equiv 1-y=1-\left[1-\left(\frac{r}{\alpha}\right)^2\right]^{1/2},
\end{equation}
\noindent where $x$ covers the range $(0,1-y_{1}]$. In terms of \eqref{newx}, equation \eqref{omegain} reads
\begin{multline}\label{important1}
x\left[2k + (2-k)x - x^2\right]\frac{d^2\varpi}{dx^2} + \left[5k + (3-5k)x - 4x^2\right]\frac{d\varpi}{dx}\\
-4(k+1)\varpi=0,
\end{multline} 
\noindent Near the origin ($x\approx 0$) $\varpi$ satisfies
\begin{equation}\label{originw}
\varpi = \left[1+\frac{4(k+1)}{5k}x\right]\varpi_{c}
\end{equation} 
\noindent where $\varpi$ is measured in the unit $\varpi_{c}$, its value at the centre, which is arbitrary. The field equations \eqref{m0int}, \eqref{h0int} take the forms
\begin{multline}\label{m0x}
\frac{dm_{0}}{dx}=\\
\alpha^{3}\frac{(1-x)\left[x(2-x)\right]^{3/2}}{(k+x)^2}\left[\frac{1}{3}x(2-x)\left(\frac{d\varpi}{dx}\right)^2+\frac{8(k+1)}{3(k+x)}\varpi^2\right],
\end{multline} 
\begin{multline}\label{p0x}
\frac{d}{dx}\delta P_{0} = -\frac{(k+1)}{(1-x)(k+x)}\delta P_{0}\\
-\left[\frac{2+(k+1)(1-x)-3(1-x)^2}{(k+x)(1-x)^2[x(2-x)]^{3/2}}\right]\alpha^{-1}m_{0}+\frac{8x(2-x)}{3(k+x)^2}\varpi\left(\frac{d\varpi}{dx}\right)\\
+\frac{\left[x(2-x)\right]^2}{3(1-x)(k+x)^2}\left(\frac{d\varpi}{dx}\right)^2-\frac{8}{3}\left[\frac{1-(k+1)(1-x)}{(k+x)^3}\right]\varpi^2,
\end{multline} 
\noindent which, near the origin, satisfy
\begin{equation}\label{originm0}
m_{0}=\left(\frac{32\sqrt{2}(k+1)}{15k^3}x^{5/2}\right)\alpha^3\varpi_{c}^2,
\end{equation}  
\begin{equation}\label{originp0}
\delta P_{0} = \left(\frac{8x}{3k^2}\right)\alpha^2\varpi_{c}^{2}.
\end{equation} 
\noindent The equations for $h_{2}, v_{2}$ as functions of $x$ now take the form
\begin{multline}\label{important4}
\frac{dv_{2}}{dx} = -\frac{2h_{2}}{k+x}\\
+\alpha^{2}\frac{2\left[x(2-x)\right]^2}{3(k+x)^3}\left[1+(k+1)(1-x)-2(1-x)^2\right]\times\\
\left[\left(\frac{d\varpi}{dx}\right)^2+\frac{4(k+1)}{x(2-x)(k+x)}\varpi^2\right],
\end{multline} 
\begin{multline}\label{important5}
\frac{dh_{2}}{dx} = \frac{(1-x)^2+(k+1)(1-x)-2}{x(2-x)(k+x)}h_{2}-\frac{2(k+x)}{[x(2-x)]^2}v_{2}\\
+\frac{\alpha^2}{3}\left\{2[x(2-x)]^2-(k+x)^2\right\}\frac{x(2-x)}{(k+x)^3}\left(\frac{d\varpi}{dx}\right)^2\\
+\frac{4\alpha^2}{3}(k+1)\left[2x^2(2-x)^2+(k+x)^2\right]\frac{\varpi^2}{(k+x)^4}.
\end{multline} 
\noindent The functions $\delta P_{0},\delta P_{2},h_{2},k_{2}$ and $v_{2}$ are measured in the unit $\alpha^2\omega_{c}^2$ and $m_{0}$ is measured in the unit $\alpha^3\omega_{c}^2$. Solutions to \eqref{important4} and \eqref{important5} can be expressed as the superposition of a particular and a complementary solution (CM.61)
\begin{equation}\label{complem}
h_{2}=h_{2}^{(p)}+\beta h_{2}^{(c)},\quad v_{2}=v_{2}^{(p)}+\beta v_{2}^{(c)}, 
\end{equation}
\noindent with $\beta$ being an integration constant. The complementary functions here satisfy the homogeneous forms of equations \eqref{important4} and \eqref{important5}
\begin{equation}\label{v2c}
\frac{dv_{2}^{(c)}}{dx}=-\frac{2h_{2}^{c}}{k+x}, 
\end{equation}
\begin{multline}\label{h2c}
\frac{dh_{2}^{(c)}}{dx}=\frac{(1-x)^2+(k+1)(1-x)-2}{x(2-x)(k+x)}h_{2}^{(c)}\\
-\frac{2(k+x)}{[x(2-x)]^2}v_{2}^{(c)}, 
\end{multline}
\noindent which have the following behaviours near the origin (CM.64)
\begin{equation}\label{h2porigin}
h_{2}^{(p)}=cx,\quad v_{2}^{(p)}=ax^2,
\end{equation}
\begin{equation}\label{h2porigin}
h_{2}^{(c)}=-kBx,\quad v_{2}^{(c)}=Bx^2
\end{equation}
\noindent where $3k^2(c-ka)=8(k+1)$ and $B$ is an arbitrary constant. On the other hand the exterior solutions \eqref{m0out},\eqref{h0out},\eqref{h2out} and \eqref{v2out} take the forms \footnote{There is a misprint in equation (CM.53), equation \eqref{extv2} here. The numerator of the second term to the right should be $(1-y_{1}^2)^{3/2}$.}
\begin{equation}\label{extm0}
m_{0} = \delta M - \frac{J^2}{r^3},\quad h_{0} = - \frac{m_{0}}{r-(1-y_{1}^2)^{3/2}},
\end{equation}
\begin{equation}\label{exth2}
h_{2} = \left[\frac{2}{(1-y_{1}^2)^{3/2}} + \frac{1}{r}\right]\frac{J^2}{r^3} + KQ_{2}^{\;2},
\end{equation}
\begin{equation}\label{extv2}
v_{2} = -\frac{J^2}{r^4} + K\frac{(1-y_{1}^2)^{3/2}}{\left\{r\left[r-(1-y_{1}^2)^{3/2}\right]\right\}^{1/2}}Q_{2}^{\;1}.
\end{equation}
\noindent At the boundary of the configuration, the interior equations \eqref{important1}-\eqref{important5} must match the exterior solutions \eqref{extm0}-\eqref{extv2}.

Finally, from \eqref{ellip}, the ellipticity of the embedded spheroid at the bounding surface takes the form
\begin{equation}\label{ellip2}
\varepsilon = \frac{3(1-x)(k+x)}{2x(2-x)}\left[h_{2}+\frac{4x(2-x)}{3(k+x)^2}\varpi(R)^2\right]-\frac{3}{2}(v_{2}-h_{2}),
\end{equation}
\noindent where $\varepsilon$ is being measured in the unit $\alpha^2\varpi_{c}^2$.

It can be observed that the structure equations preserve the same form as considered in \cite{chandra1974}. The only significant change corresponds to the modulus condition \eqref{modulok} which, we emphasize, is crucial for the strict analysis of the regime $R_{s}<R<(9/8)R_{s}$.
\section{Results}
\label{sect5}
In this section we present the results of integrations of the Hartle structure equations and derived surface properties for a slowly rotating Schwarzschild star with negative pressure, in the unstudied regime $R_{s}<R<(9/8)R_{s}$. We followed similar methods to those used by \cite{chandra1974}, in particular, units were chosen such that derived quantities are dimensionless. Similarly we constructed several configurations under quasi-stationary contraction, by varying the radius of the star through the parameter $R/R_{s}$. For convenience we introduce the `Schwarzschild deviation parameter' $\zeta\equiv(R-R_{s})/R_{s}$.

The integrations were performed numerically using the Runge-Kutta-Fehlberg (RKF) adaptive method in Python 3.4 \citep{kiusalaas2010,newman2012}. It is well known that adaptive methods are usually more convenient, than the standard fourth-order Runge-Kutta, when the function to be integrated changes rapidly near some point. In such situations, setting a constant step of integration $h$ on the whole integration range might not be appropriate and we are forced to adjust the step to maintain the truncation error within prescribed limits. In our particular case, we found that the (RKF) method provided a fast, reliable and stable technique to integrate the structure equations near and below the Buchdahl radius.

In our routine the condition \eqref{modulok} was specified which is key to analyze the region below the Buchdahl radius. We have checked our code by reproducing the computations found in \citep{chandra1974} for $R\geq 1.125R_{s}$. We found agreement up to the fourth decimal place in some cases (see Table \ref{table1}). In contrast to the CM paper, we are measuring the mass quadrupole moment $Q$ in units of $J^2/M_{0}$. It is useful to introduce the quantity \citep{bradley2009} 
\begin{equation}\label{Q}
\frac{\Delta Q}{Q}\equiv\frac{Q-Q_{kerr}}{Q_{kerr}}  
\end{equation}
\begin{center}
\begin{table*}
\label{tab:table1}
\caption{Integral and surface properties of a slowly rotating `Schwarzschild star' for several values of the deviation parameter $\zeta\equiv\frac{R-R_{s}}{R_{s}}$, where $R$ is the radius of the star and $R_{s}=2M_{0}$ is the Schwarzschild radius. We use geometrized units ($c=G=1$). The angular velocity relative to the local ZAMO $\varpi(R)= (\Omega - \omega)\vert_{r=R}$ is given in units of $J/R_{s}^3$. The moment of inertia $I$ is in the unit $R_{s}^3$. The ratio $\delta M^{H}/M$ denotes the original Hartle's fractional change in mass, as given by \eqref{m0out}, measured in units of $J^2/R_{s}^4$. The ratio $\delta M/M$ corresponds to the amended fractional change of mass as given by \eqref{deltamc}. The ratio $\Delta Q/Q$ defined in \eqref{Q} corresponds to the relative deviation of the mass quadrupole moment from that of the Kerr metric. We measure the quadrupole moment $Q$ in units of $J^2/M_{0}$ so the Kerr factor $\bar{q}=QM_{0}/J^2$ corresponds to the unity. The ellipticity $\varepsilon$ is measured in units of $J^2/M^4$. All the quantities are computed at the surface of the configuration.
The digit in parenthesis following each entry corresponds to the power of ten by which the entry is multiplied.}
\hfill{}
\resizebox{\textwidth}{!}{
\begin{tabular}{ccccccccc}
\hline\hline 
$\zeta$ & $\varpi(R)$  & $\Omega$ & $I$ & $I_{N}=I/M_{0}R^2$ & $\delta{M^{H}}/M$ & $\delta M/M$ & $\Delta Q/Q$ & $\varepsilon$  \\
\hline
99.0 & 4.958538 (-4) & 4.978538 (-4) & 2.008621 (+3) & 4.017243 (-1) & 9.950087 (-4) & 4.901465 (-1) & 6.093865 (+2) & 3.848749 (-2)\\
49.0 & 1.966802 (-3) & 1.982802 (-3) & 5.043366 (+2) & 4.034693 (-1) & 3.960078 (-3) & 9.608127 (-1) & 2.970315 (+2) & 7.583446 (-2)\\
34.0 & 3.984776 (-3) & 4.031423 (-3) & 2.480513 (+2) & 4.049817 (-1) & 8.046649 (-3) & 1.348994 (0)  & 2.033647 (+2) & 1.069368 (-1)\\
9.0  & 4.582047 (-2) & 4.782047 (-2) & 2.091154 (+1) & 4.182308 (-1) & 9.493450 (-2) & 4.065297 (0)  & 4.794558 (+1) & 3.345472 (-1)\\
7.0  & 6.994179 (-2) & 7.384804 (-2) & 1.354132 (+1) & 4.231662 (-1) & 1.463199 (-1) & 4.811075 (0)  & 3.575002 (+1) & 4.014170 (-1)\\
4.0  & 1.662453 (-1) & 1.822453 (-1) & 5.487109 (0)  & 4.389687 (-1) & 3.588505 (-1) & 6.482244 (0)  & 1.786438 (+1) & 5.643897 (-1)\\
3.0  & 2.462305 (-1) & 2.774805 (-1) & 3.603856 (0)  & 4.504820 (-1) & 5.439746 (-1) & 7.172851 (0)  & 1.217989 (+1) & 6.437757 (-1)\\
2.0  & 3.969839 (-1) & 4.710580 (-1) & 2.122880 (0)  & 4.717512 (-1) & 9.162061 (-1) & 7.689061 (0)  & 6.837185 (0)  & 7.287112 (-1)\\
1.0  & 7.029353 (-1) & 9.529353 (-1) & 1.049389 (0)  & 5.246945 (-1) & 1.819485 (0)  & 6.951236 (0)  & 2.263128 (0)  & 7.473735 (-1)\\
0.50 & 8.871430 (-1) & 1.479735 (0)  & 6.757963 (-1) & 6.007078 (-1) & 2.756967 (0)  & 4.905143 (0)  & 6.358729 (-1) & 6.183357 (-1)\\
0.40 & 8.917486 (-1) & 1.620611 (0)  & 6.170509 (-1) & 6.296438 (-1) & 2.992654 (0)  & 4.246032 (0)  & 4.065443 (-1) & 5.646304 (-1)\\
0.30 & 8.574163 (-1) & 1.767748 (0)  & 5.656913 (-1) & 6.694571 (-1) & 3.224895 (0)  & 3.506898 (0)  & 2.204677 (-1) & 4.981916 (-1)\\
0.20 & 7.481799 (-1) & 1.905587 (0)  & 5.247725 (-1) & 7.288508 (-1) & 3.412176 (0)  & 2.725004 (0)  & 8.451602 (-2) & 4.205147 (-1)\\
0.15 & 6.439067 (-1) & 1.958939 (0)  & 5.104803 (-1) & 7.719929 (-1) & 3.454114 (0)  & 2.347975 (0)  & 3.810123 (-2) & 3.806335 (-1)\\
0.125& 5.727118 (-1) & 1.977375 (0)  & 5.057207 (-1) & 7.991636 (-1) & 3.442297 (0)  & 2.176744 (0)  & 1.155989 (-2) & 3.457449 (-1)\\
\hline\hline
0.120 & 5.796710 (-1) & 2.003231 (0) & 4.991934 (-1) & 7.959078 (-1) & 3.318006 (0) & 2.239330 (0) & 1.498300 (-2) & 3.570681 (-1)\\
0.115 & 5.912024 (-1) & 2.034000 (0) & 4.916420 (-1) & 7.909140 (-1) & 3.146362 (0) & 2.206228 (0) & 8.225660 (-3) & 3.498988 (-1)\\
0.110 & 5.988341 (-1) & 2.061216 (0) & 4.851503 (-1) & 7.875177 (-1) & 3.013208 (0) & 2.207648 (0) & 3.292989 (-3) & 3.448637 (-1)\\
0.10  & 6.105343 (-1) & 2.113163 (0) & 4.732240 (-1) & 7.821885 (-1) & 2.775629 (0) & 2.191611 (0) & 4.029515 (-3) & 3.372674 (-1)\\
5.0 (-2) & 5.749893 (-1) & 2.302664 (0) & 4.342794 (-1) & 7.878086 (-1) & 2.153885 (0) & 2.105008 (0) & 8.882128 (-3) & 3.476065 (-1)\\
1.0 (-2) & 3.175043 (-1) & 2.258684 (0) & 4.427355 (-1) & 8.680238 (-1) & 1.984771 (0) & 1.983793 (0) & 1.475649 (-3) & 3.873924 (-1)\\
5.0 (-3) & 2.331664 (-1) & 2.203463 (0) & 4.538308 (-1) & 8.986527 (-1) & 1.990785 (0) & 1.990653 (0) & 6.363087 (-4) & 3.904995 (-1)\\
1.0 (-3) & 1.091218 (-1) & 2.103133 (0) & 4.754809 (-1) & 9.490627 (-1) & 1.997866 (0) & 1.997866 (0) & 1.001120 (-4) & 3.863623 (-1)\\
5.0 (-4) & 7.045530 (-2) & 2.067458 (0) & 4.836856 (-1) & 9.664047 (-1) & 1.997955 (0) & 1.997943 (0) & 5.726721 (-5) & 3.832646 (-1)\\ 
1.0 (-4) & 3.109358 (-2) & 2.030493 (0) & 4.924910 (-1) & 9.847851 (-1) & 1.999557 (0) & 1.999555 (0) & 1.096807 (-6) & 3.791711 (-1)\\
5.0 (-6) & 6.929922 (-3) & 2.006899 (0) & 4.982809 (-1) & 9.965519 (-1) & 1.999977 (0) & 1.999977 (0) & 5.326670 (-7) & 3.760089 (-1)\\   
1.0 (-6) & 3.097805 (-3) & 2.003091 (0) & 4.992282 (-1) & 9.984544 (-1) & 1.999995 (0) & 1.999995 (0) & 1.060549 (-7) & 3.754575 (-1)\\
5.0 (-8) & 6.925427 (-4) & 2.000692 (0) & 4.998269 (-1) & 9.996538 (-1) & 1.999999 (0) & 1.999999 (0) & 5.287854 (-9) & 3.751032 (-1)\\
1.0 (-12)& 3.097091 (-6) & 2.000003 (0) & 4.999992 (-1) & 9.999984 (-1) & 2.0 (0)      & 2.0 (0)      & 1.056932 (-13)& 3.750004 (-1)\\
1.0 (-14)& 3.095714 (-7) & 2.0 (0)      & 4.999999 (-1) & 9.999998 (-1) & 2.0 (0)      & 2.0 (0)      & 1.110223 (-15)& 3.750000 (-1)\\
\hline\hline
\end{tabular}}
\hfill{}
\label{table1}
\end{table*}
\end{center}
\noindent which corresponds to the relative deviation of the quadrupole moment from the Kerr metric value ($Q_{kerr}=J^2/M_{0}$). 

One important observation by \cite{chandra1974} is that the structure equations can actually be integrated at the Buchdahl radius, i.e., $y_{1}=1/3$ and $k=0$, by considering the expansions (CM.69)-(CM.71). We reproduced these results (see Table~\ref{table1}) with very good agreement, except for the mass quadrupole moment where we found $Q=2.02311$ (in units of $J^2/R_{s}$). The fact that the structure equations are integrable at $R=(9/8)R_{s}$ might be seen as an early indication that the region below this limit is potentially interesting. This hypothesis has been investigated in this paper inspired by the results of \cite{mazur2015}. In the following, we present some plots of our results and further analysis.  

\begin{figure}
\includegraphics[scale=0.4]{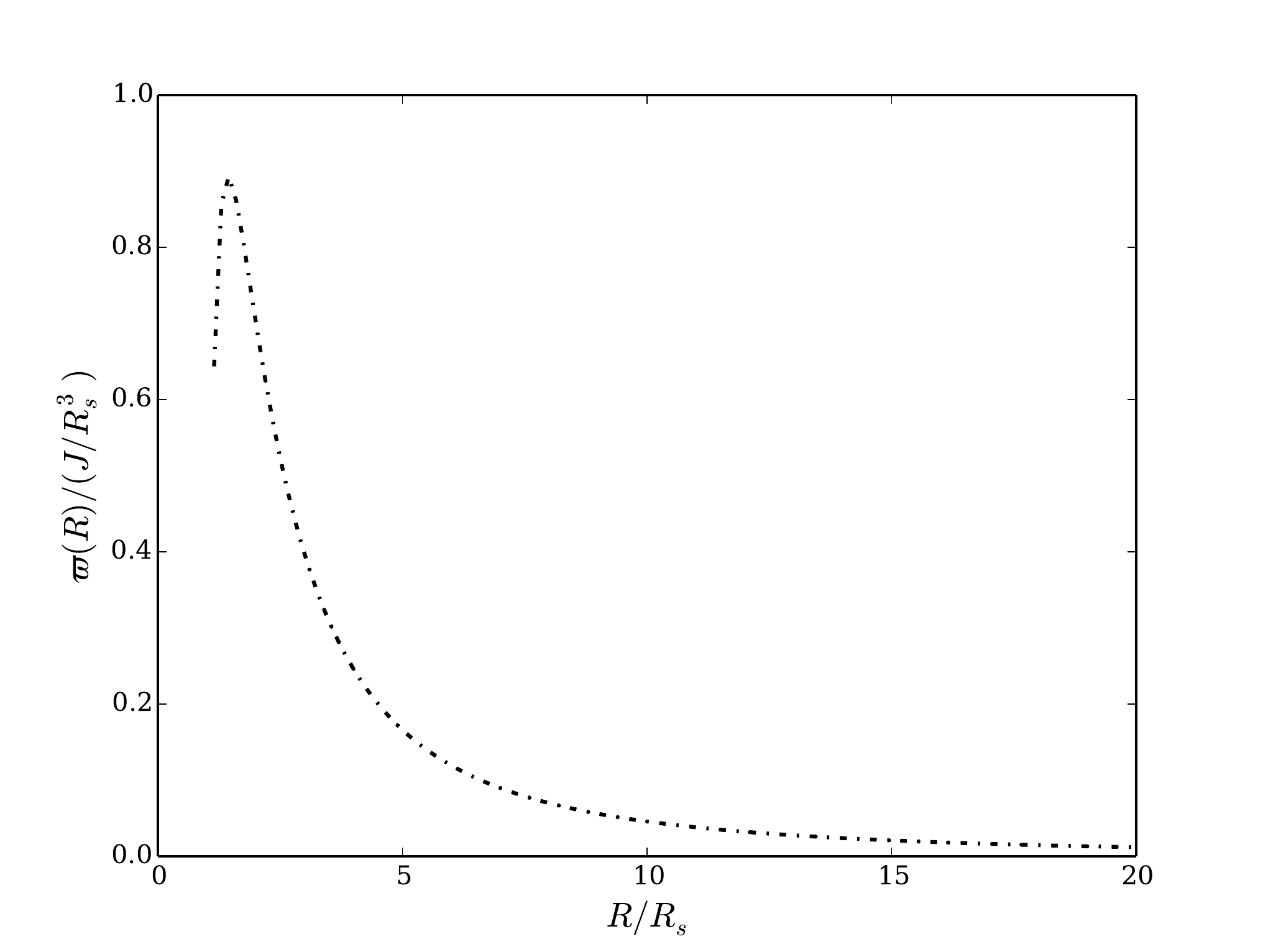}
\caption{The angular velocity $\varpi=(\Omega-\omega)|_{r=R}$ (in units of $J/R_{s}^3$) relative to the local ZAMO, plotted as a function of the compactness parameter $R/R_{s}$ above the Buchdahl-Bondi bound $R>1.125R_{s}$.}
\label{fig4}
\end{figure}

In Fig.~\ref{fig4} we plot the surface value of the fluid angular velocity relative to the local ZAMO $\varpi(R)$, versus the `compactness parameter' $R/R_{s}$, above the Buchdahl bound. Notice that $\varpi(R)$ reaches a maximum near to $R=1.4 R_{s}$ and then approaches zero in the Newtonian limit $R\to\infty$. These results are in very good agreement with \citep{chandra1974}.
\begin{figure}
\centering
\includegraphics[scale=0.4]{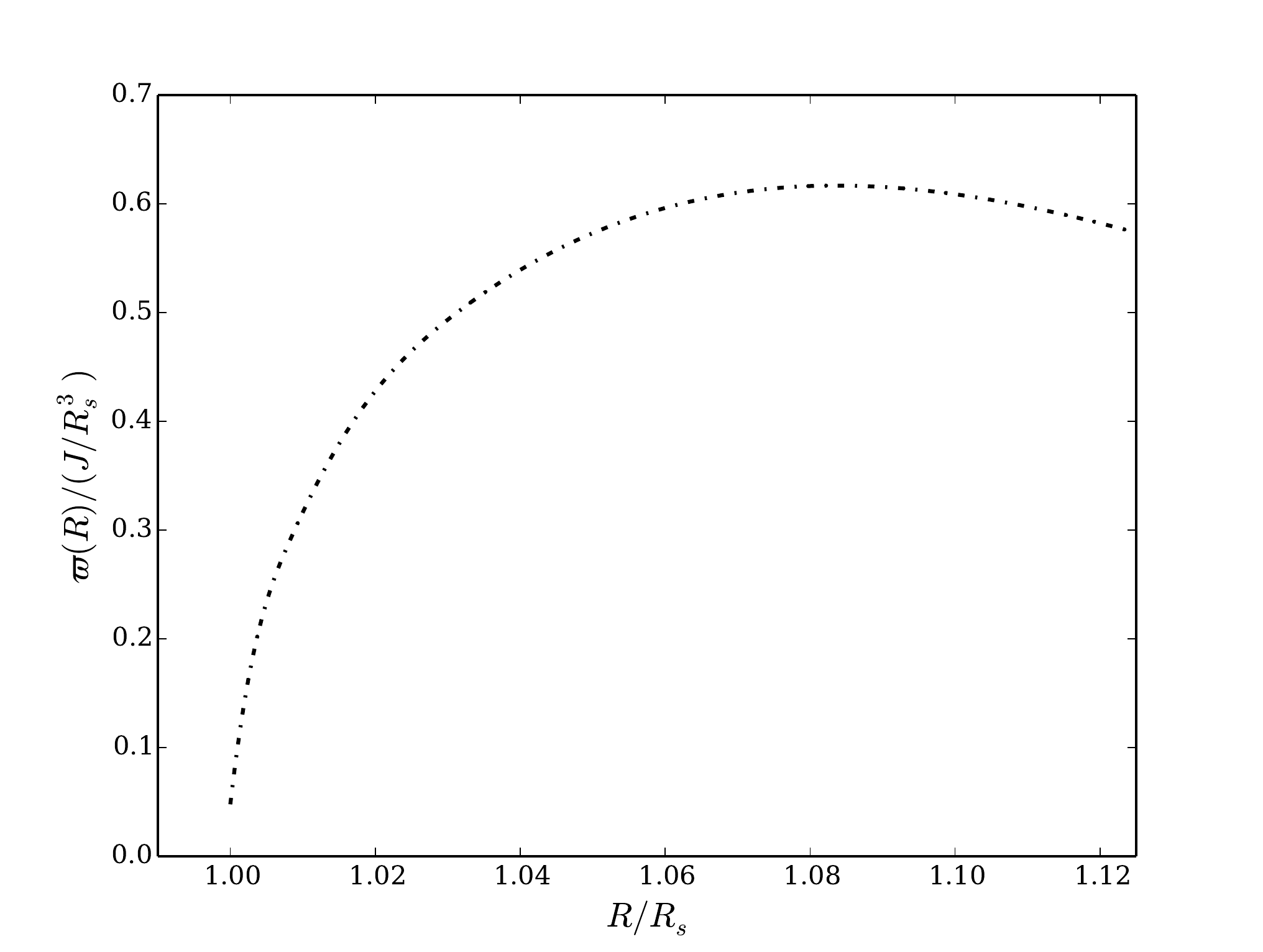}
\caption{The angular velocity $\varpi=(\Omega-\omega)|_{r=R}$ (in units of $J/R_{s}^3$) relative to the local ZAMO, plotted as a function of the compactness parameter $R/R_{s}$ in the region $R_{s}<R<1.125R_{s}$.}
\label{fig5}
\end{figure}
\begin{figure}
\centering
\includegraphics[scale=0.4]{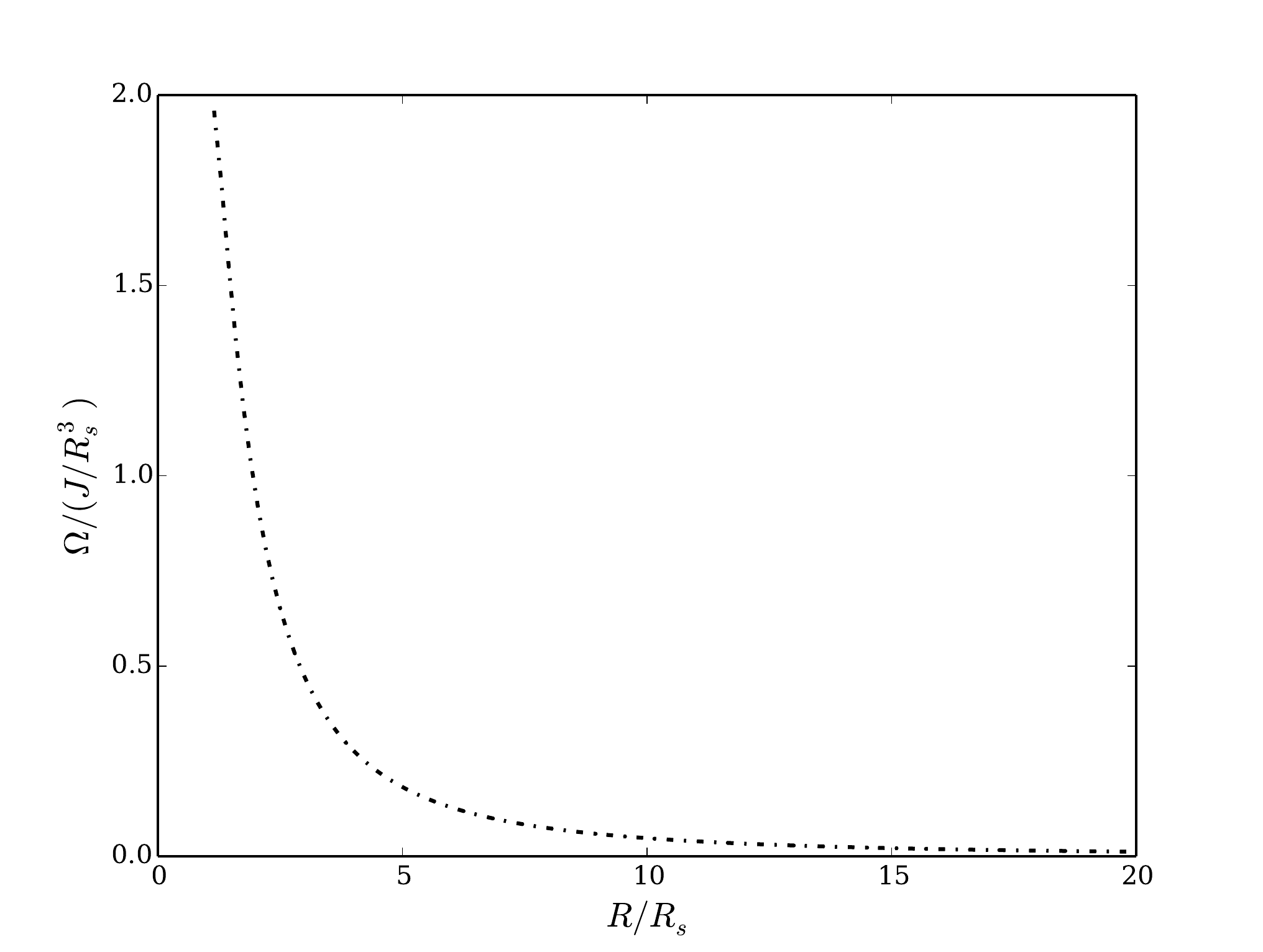}
\caption{The angular velocity $\Omega$ (in units of $J/R_{s}^3$) relative to an observer at infinity, plotted as a function of the compactness parameter $R/R_{s}$ above the Buchdahl bound.}
\label{fig6}
\end{figure}
\begin{figure}
\centering
\includegraphics[scale=0.4]{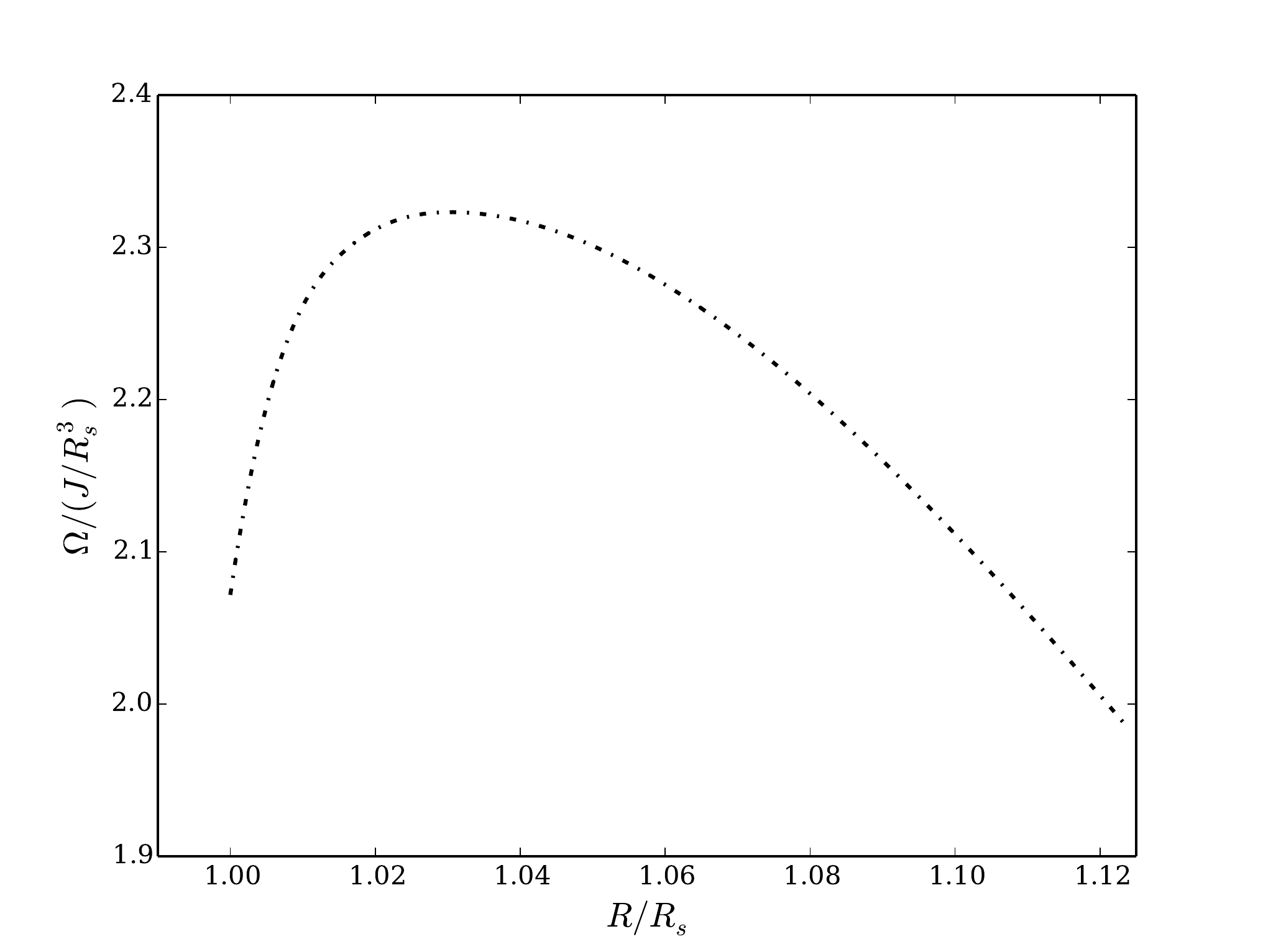}
\caption{The angular velocity $\Omega$ (in units of $J/R_{s}^3$) plotted as a function of the compactness parameter $R/R_{s}$ in the region $R_{s}<R<1.125R_{s}$.}
\label{fig7}
\end{figure}

Figure~\ref{fig5} shows the angular velocity $\varpi(R)$ relative to the local ZAMO as a function of $R/R_{s}$ in the regime $R_{s}<R<(9/8)R_{s}$. It is observed that in the limit when the radius of the star $R$ approaches the Schwarzschild radius $R_{s}$, $\varpi$ tends to zero. In connection with this, Fig.~\ref{fig7} shows the angular velocity $\Omega$, relative to a distant observer, as a function of $R/R_{s}$ in the region $R_{s}<R<(9/8)R_{s}$. Notice the increase in the angular velocity reaching a maximum at $R/R_{s}\approx 1.03$, and the subsequent decrease tending to the value $2$ (in units of $J/R_{s}^3$) when $R\to R_{s}^{+}$. In this limit, $\varpi\to 0$ and the angular velocity $\Omega=\omega$ is a constant indicating a \emph{rigidly rotating body} with no differential surface rotation \citep{marsh2014}.

It can be shown that the value $\Omega=\omega=2$ (in units of $J/R_{s}^3$) for the angular velocity of the super-compact Schwarzschild star in the gravastar limit ($\zeta\sim 10^{-14}$) is consistent with that of the Kerr black hole limit. It is well known that in the Kerr spacetime, a radially falling test particle with zero angular momentum acquires an angular velocity when it approaches the spinning black hole. The angular velocity as measured by a distant ZAMO is given by
\begin{equation}\label{angular}      
\omega=\frac{d\phi}{dt}=\frac{2aM_{0}r}{(r^2+a^2)^2-\Delta(r)a^2\sin^2\theta},
\end{equation}
\noindent where $a\equiv J/M_{0}$ and $\Delta(r)\equiv r^2-2M_{0}r+a^2$. Notice that positive $a$ implies positive $\omega$, therefore the particle will rotate in the spinning direction of the black hole. This is the so-called dragging effect in Kerr geometry. At the `event horizon' \eqref{angular} satisfies
\begin{equation}\label{angular2}      
\omega_{{\tiny bh}}=\frac{a}{2M_{0}r_{+}},
\end{equation}
\noindent where $r_{+}=M_{0}+(M_{0}^2-a^2)^{1/2}$. Equation \eqref{angular2} corresponds to the angular velocity of the Kerr black hole. In the slowly rotating approximation $(\xi\equiv a/M_{0}<<1)$ a straightforward calculation from \eqref{angular2} shows that
\begin{equation}\label{angular3}      
\Omega=\omega_{bh}\approx\frac{a}{4M_{0}^2}+\mathcal{O}(\xi^2)=2\left(\frac{J}{R_{s}^3}\right)+\mathcal{O}(\xi^2)
\end{equation}
\noindent which is consistent with our numerical results for $\Omega$ in the gravastar limit $\zeta\sim 10^{-14}$ (see Table~\ref{table1}).

\begin{figure}
\centering
\includegraphics[scale=0.4]{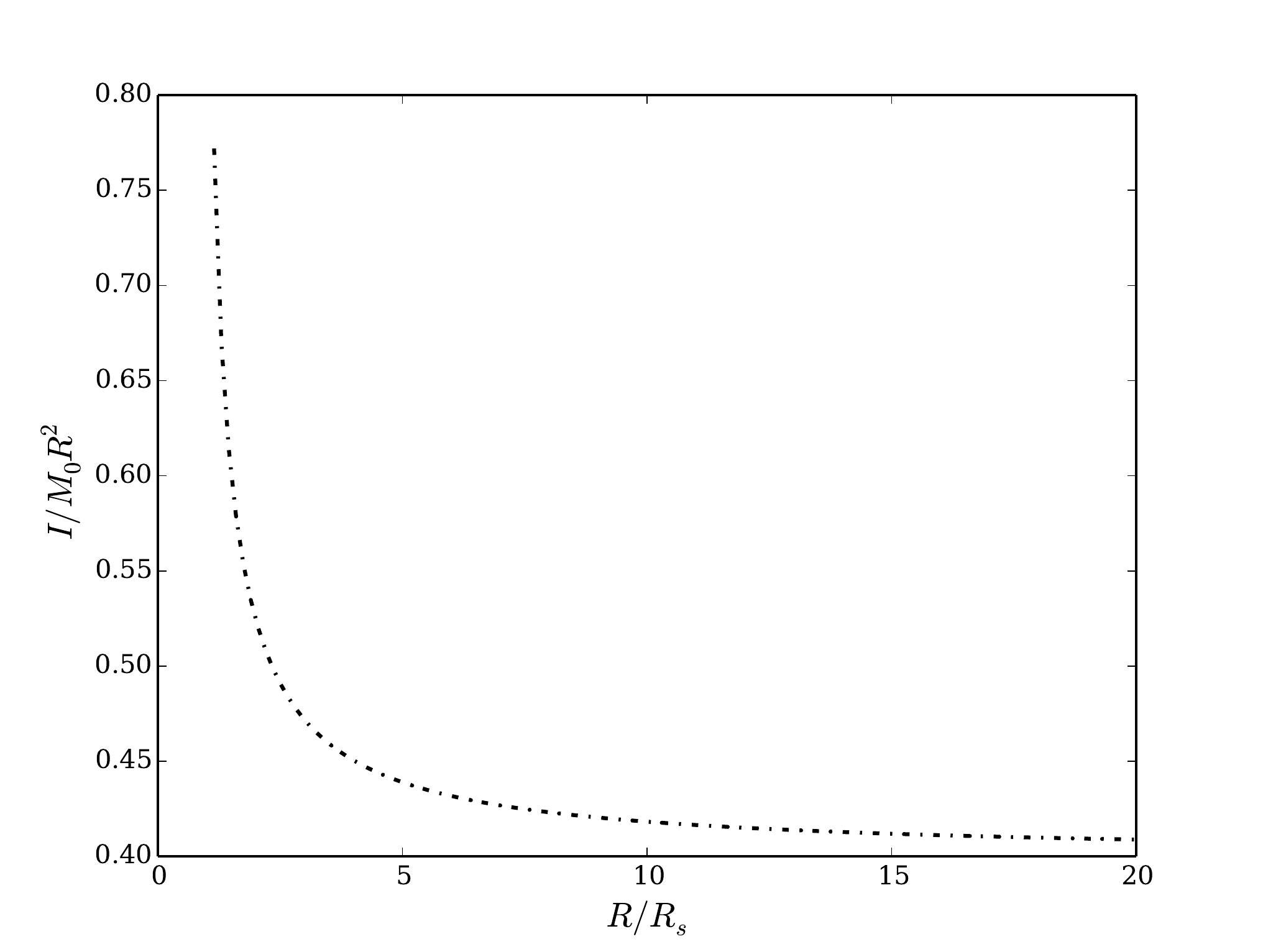}
\caption{The normalized moment of inertia $I_{N}$ plotted as a function of the compactness parameter $R/R_{s}$ above the Buchdahl bound.}
\label{fig8}
\end{figure}

From Fig.~\ref{fig8} it can be observed that the normalized moment of inertia approaches the value 0.8 at the Buchdahl bound. For large values of $R$ we notice that $I_{N}$ tends to the value 0.4 which is the well known moment of inertia of a sphere in consistency with the Newtonian limit. 

\begin{figure}
\centering
\includegraphics[scale=0.4]{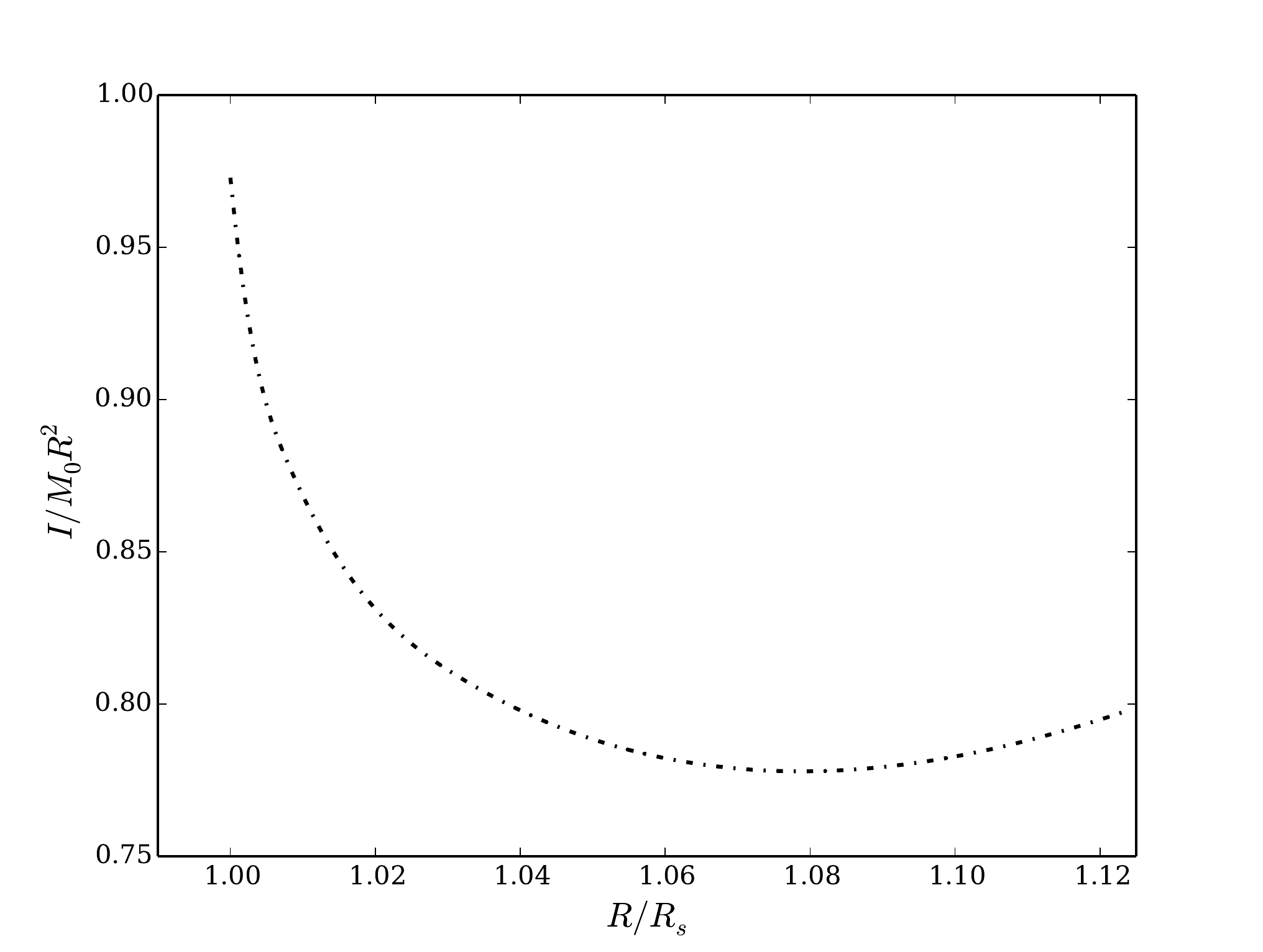}
\caption{The normalized moment of inertia $I/M_{0}R^2$ plotted as a function of the compactness parameter $R/R_{s}$ in the region $R_{s}<R<1.125 R_{s}$. Notice the approach to 1 of the moment of inertia (normalized) in the gravastar limit $R\to R_{s}^{+}$.}
\label{fig9}
\end{figure}

Figure~\ref{fig9} shows the normalized moment of inertia versus the factor $R/R_{s}$ in the region $R_{s}<R<(9/8)R_{s}$. Notice how $I_{N}$ approaches 1 systematically when $R\to R_{s}^{+}$. This is in remarkable agreement with the Kerr value in the slowly rotating approximation which is given by
\begin{equation}\label{inertiakerr}      
I=\frac{J}{\omega_{bh}}\approx 4M_{0}^3+O(\xi^2).
\end{equation}
\indent 

\begin{figure}
\centering
\includegraphics[scale=0.4]{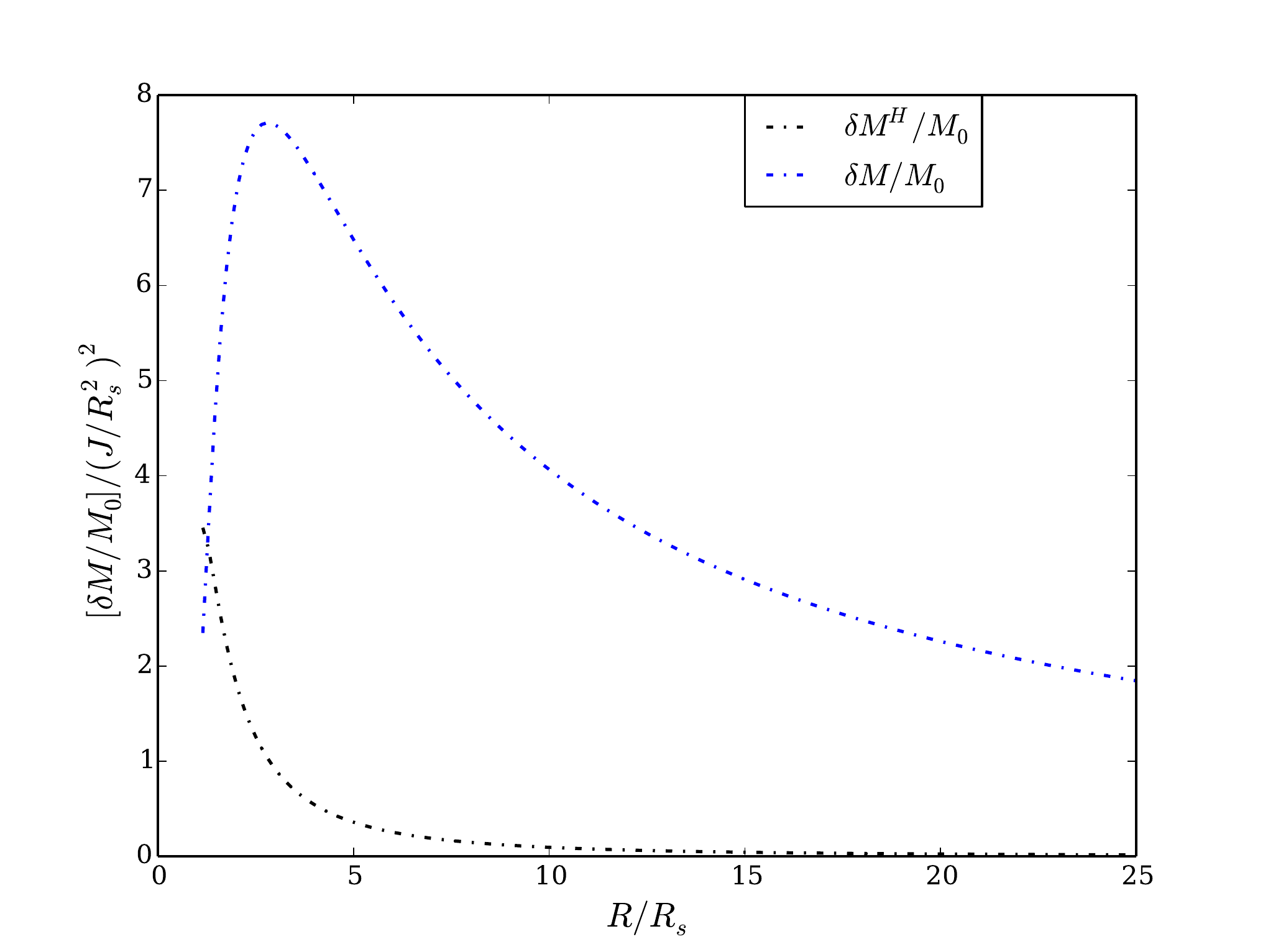}
\caption{The original $\delta M^{H}/M_{0}$ and amended $\delta M/M_{0}$ fractional change of mass against the compactness parameter $R/R_{s}$ above the Buchdahl bound.}
\label{fig10}
\end{figure}

In Fig.~\ref{fig10} the original \eqref{m0out} and amended \eqref{deltamc} change of mass are plotted as a function of the compactness parameter $R/R_{s}$, for $R\geq(9/8)R_{s}$. Notice that $\delta M/M_{0}$ reaches a maximum at $R/R_{s}\sim 2.81$ and then decreases in the Newtonian limit. Notice that this decrease is slower than the one obtained from the original \eqref{m0out}. These results are in very good agreement with \citep{reina2016}. 

\begin{figure}
\centering
\includegraphics[scale=0.4]{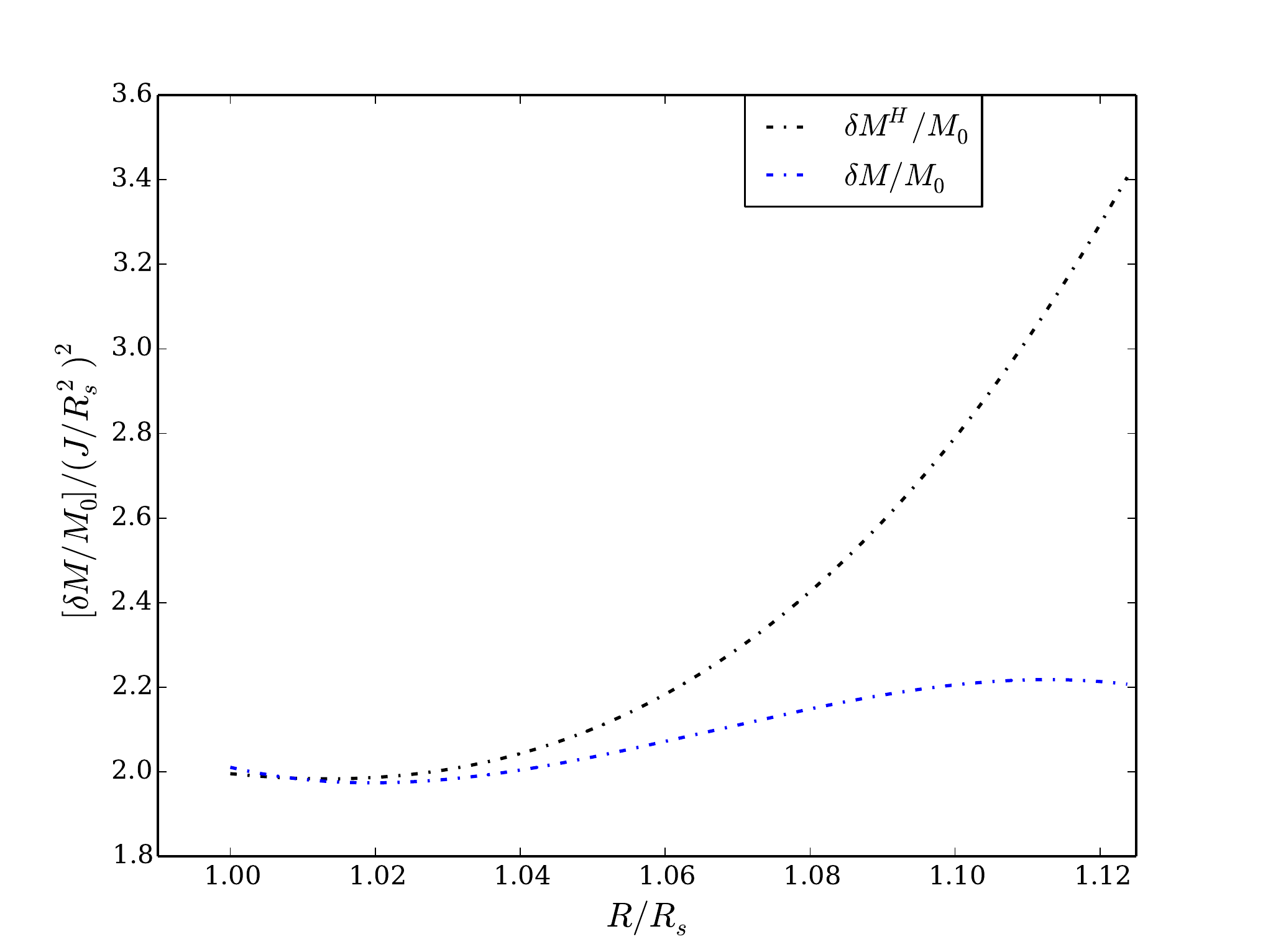}
\caption{The original $\delta M^{H}/M_{0}$ and amended $\delta M/M_{0}$ fractional change of mass as a function of the compactness parameter $R/R_{s}$ in the region $R_{s} < R < 1.125R_{s}$.}
\label{fig11}
\end{figure}

Figure \ref{fig11} shows the original and amended fractional change in mass, as a function of the parameter $R/R_{s}$, for $R_{s}<R<(9/8)R_{s}$. Notice the systematic decrease of $\delta M^{H}/M_{0}$ and $\delta M/M_{0}$ for $R$ below the Buchdahl bound, and their subsequent approach to the value 2 in the gravastar limit $R\to R_{s}^{+}$. Notice that in this limit the additional term for the change of mass in \eqref{deltamc} is negligible.

\begin{figure}
\centering
\includegraphics[scale=0.4]{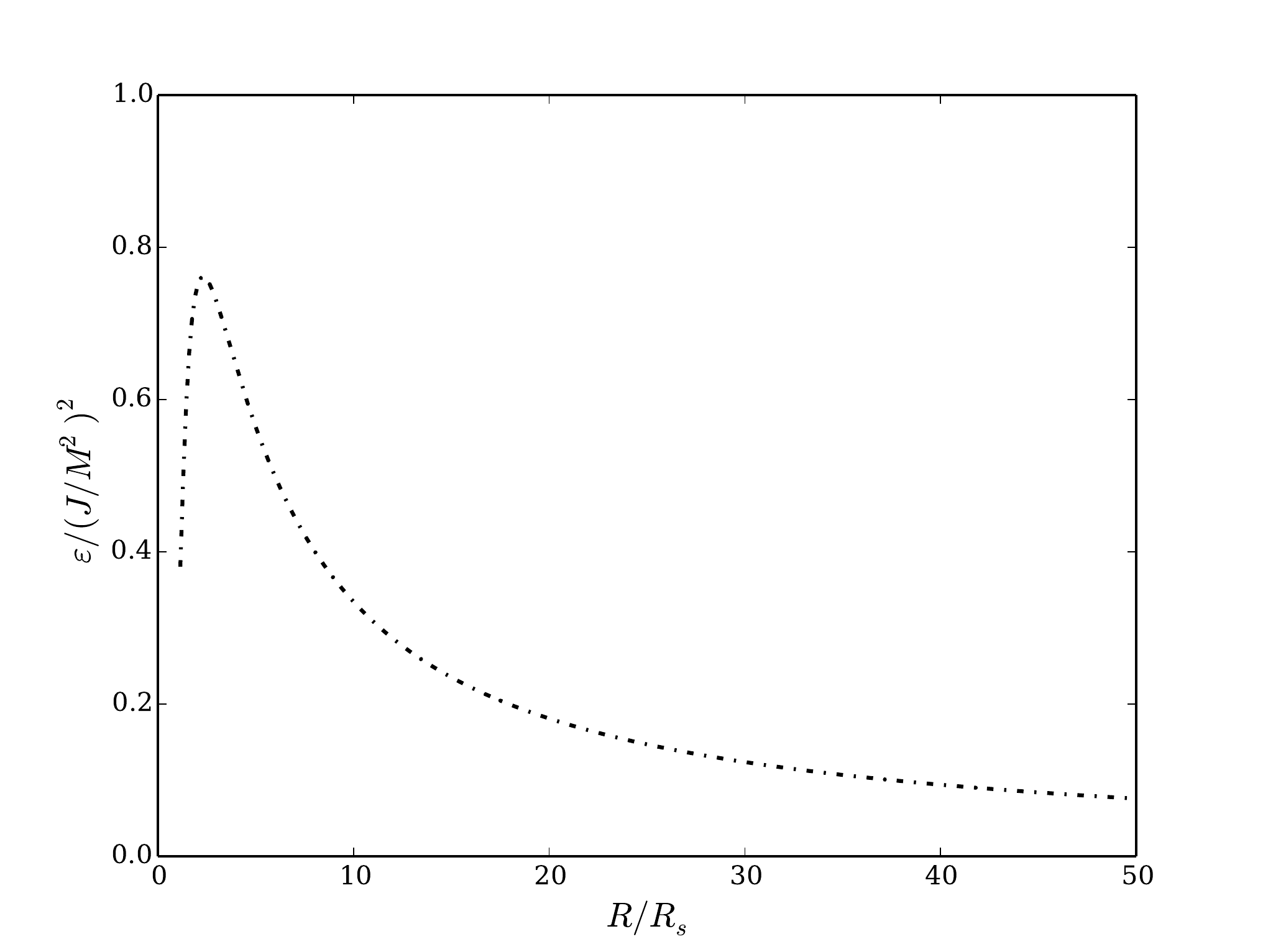}
\caption{The ellipticity of the bounding surface (in units of $J^2/M^4$) as a function of the compactness parameter $R/R_{s}$ above the Buchdahl bound.}
\label{fig12}
\end{figure}

\begin{figure}
\centering
\includegraphics[scale=0.4]{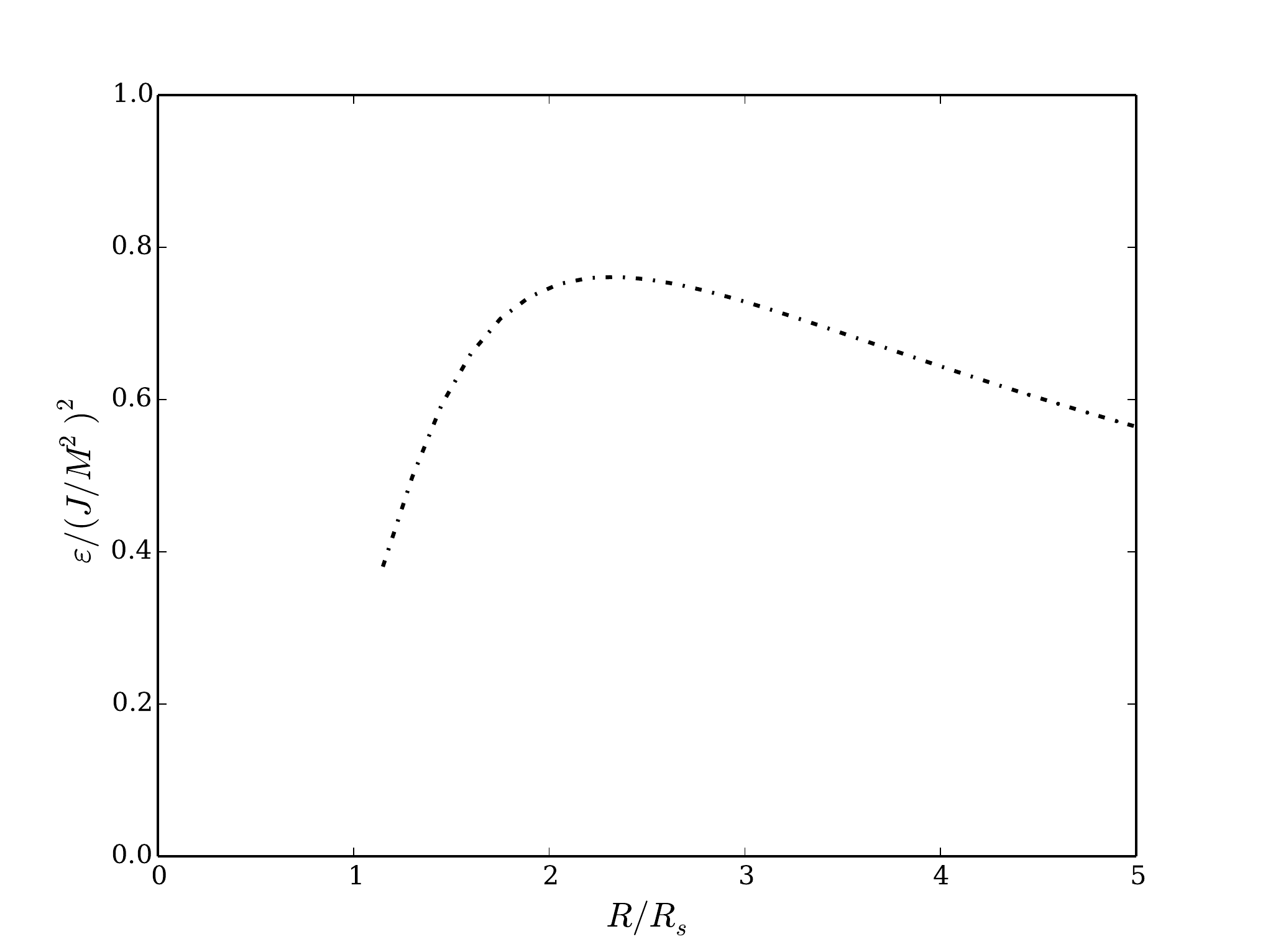}
\caption{The ellipticity of the bounding surface (in units of $J^2/M^4$) as a function of the compactness parameter $R/R_{s}$ above the Buchdahl limit. The horizontal axis has been plotted with higher resolution to show more detail.}
\label{fig13}
\end{figure}

Figures \ref{fig12} and \ref{fig13} show the ellipticity $\varepsilon(R)$ of the bounding surface, as defined in \eqref{ellip2}, plotted as a function of the compactness parameter $R/R_{s}$ above the Buchdahl bound, for a Schwarzschild star with fixed total mass and angular momentum. Notice the non-monotonic behaviour reaching a maximum at $R/R_{s}\sim 2.4$. In principle, under adiabatic contraction, the star would be flattened as expected due to the rotation. However, a peculiar behaviour occurs below the maximum $R/R_{s}\sim 2.4$ where $\varepsilon(R)$ decreases indicating that the configuration becomes more spherical. These results are in very good agreement with \citep{chandra1974,miller1977,abramowicz1990}.

\begin{figure}
\centering
\includegraphics[scale=0.4]{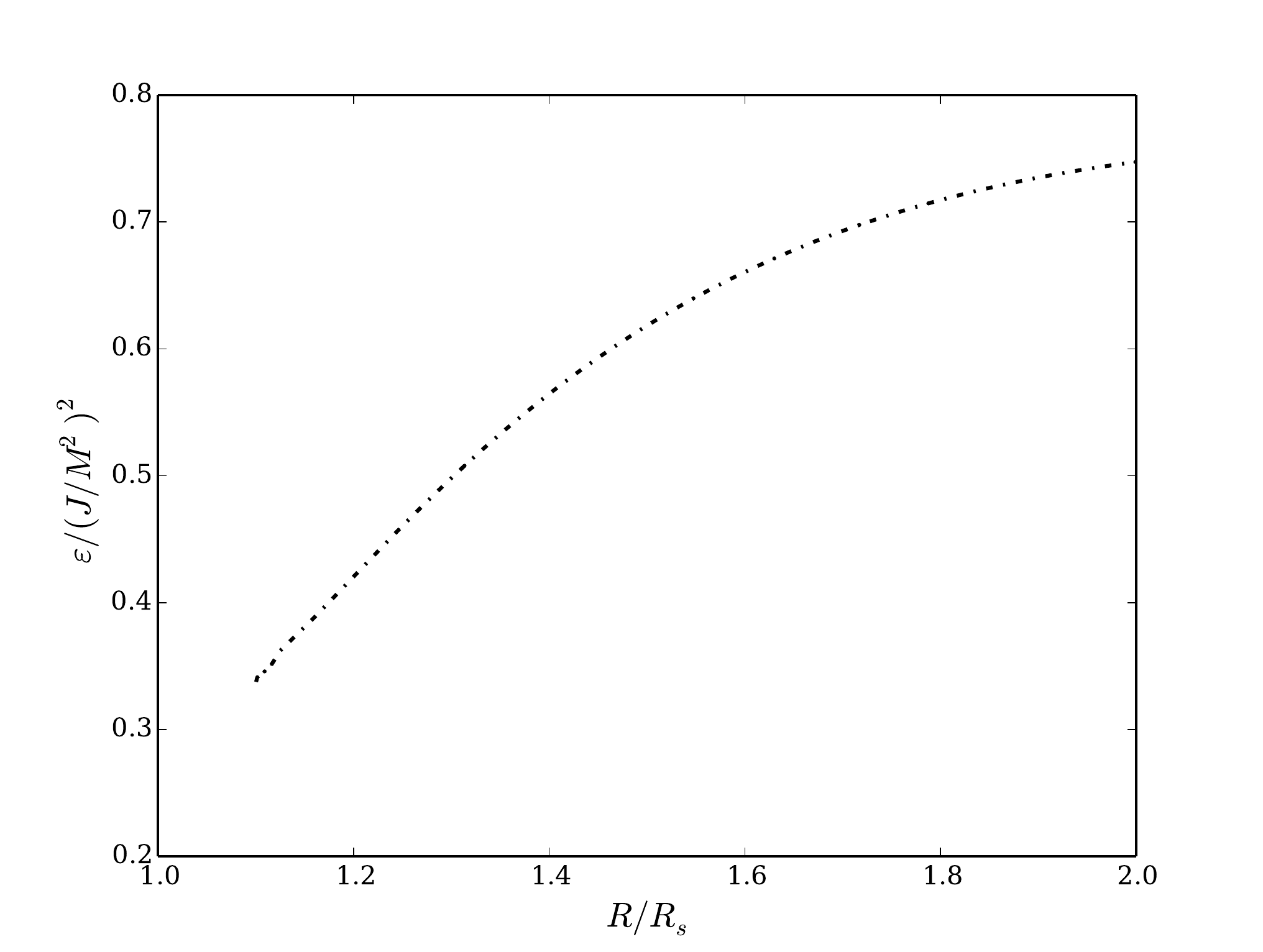}
\caption{The ellipticity of the surface (in units of $J^2/M^4$) plotted against $R/R_{s}$ in the regime $1.10<R/R_{s}<2$.}
\label{fig14}
\end{figure}

The reversal of ellipticity for a slowly rotating relativistic star, has been the subject of lively discussion in the literature \citep{chandra1974,miller1977,abramowicz1990,abramowicz1993,chakra1992}. In connection with this, in Fig.~\ref{fig14} we plot the ellipticity as a function of $R/R_{s}$ in the region $1.10<R/R_{s}<2$. Notice the continuous decrease of the ellipticity as the compactness increases. In the limit when $R\to R_{s}^{+}$ the eccentricity tends to the value 0.375 (in units of $J^2/M^4$). Notice that in the relativistic case, in contrast to the Newtonian approximation, the eccentricity is a much more complicated quantity which depends not only of the centrifugal effects (as determined by $\varpi$) but also of the behaviour of the functions $h_{2}(y_{1})$ and $v_{2}(y_{1})$.

\begin{figure}
\centering
\includegraphics[scale=0.4]{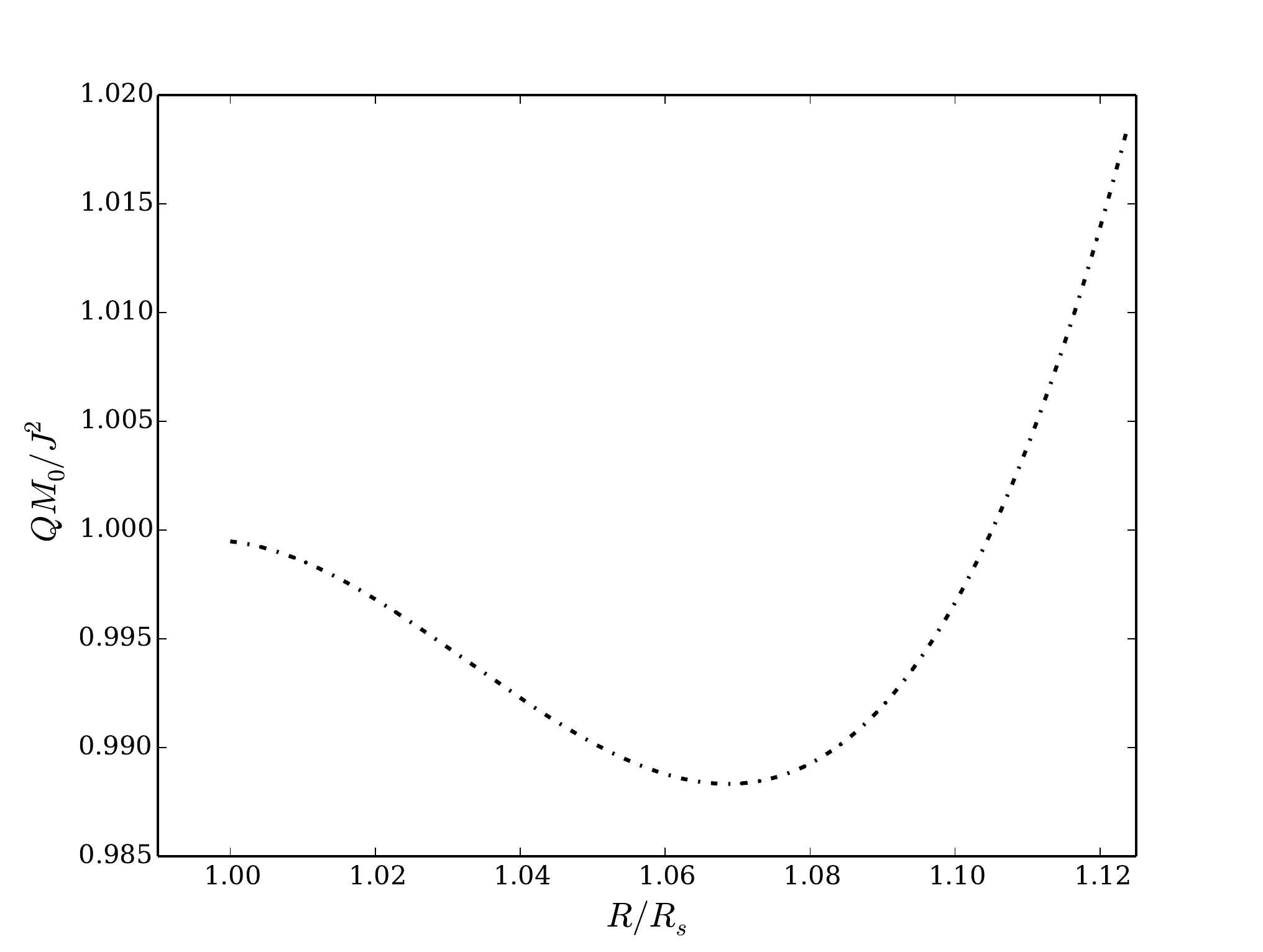}
\caption{The Kerr factor $\bar{q}=QM_{0}/J^2$ plotted as a function of the compactness parameter $R/R_{s}$ in the regime $R_{s}<R<(9/8)R_{s}$. Notice the approach to the Kerr value $\bar{q}=1$ in the gravastar limit $R\to R_{s}^{+}$. In Table~\ref{table1} it is shown that the relative deviation $\Delta Q/Q$ is of the order of $10^{-15}$ for $\zeta\sim 10^{-14}$.  }
\label{fig15}
\end{figure}

Finally in Fig.~\ref{fig15} the Kerr factor $\bar{q}=QM_{0}/J^2$ \citep{thorne1971,miller1977} is plotted as a function of the parameter $R/R_{s}$. Notice the approach to the Kerr value $\bar{q}=1$ when $R\to R_{s}^{+}$. A remarkable and unprecedented result is that relative deviations of the mass quadrupole moment as given by \eqref{Q} are of the order of $10^{-15}$ in the gravastar limit $R\to R_{s}^{+}$ with $\zeta\sim 10^{-14}$. Therefore, we conclude that the exterior metric to a slowly rotating super-compact Schwarzschild star (with negative pressure) in the gravastar limit $R\to R_{s}^{+}$, agrees to an accuracy of 1 part in $10^{15}$ with the Kerr metric.
\section{Concluding Remarks}
\label{sect6}
Motivated by recent investigations of \cite{mazur2015} and the methods introduced by \cite{hartle1967} and \cite{chandra1974} in the study of slowly rotating relativistic masses, we have presented in this paper results for integral and surface properties of a slowly rotating super-compact Schwarzschild star in the unstudied regime $R_{s}<R<(9/8)R_{s}$. We found that the angular velocity $\varpi$ relative to the local ZAMO tends to zero in the gravastar limit $R\to R_{s}^{+}$. This result indicates that the super-compact Schwarzschild star rotates rigidly with no differential surface rotation. Furthermore the angular velocity $\Omega$ of the super-compact Schwarzschild star, in the gravastar limit, is constant and approaches the corresponding Kerr value in the slowly rotating approximation.

Additionally, we found that the normalized moment of inertia $I/M_{0}R^2$ approaches 1 systematically when $R\to R_{s}^{+}$. This result is in agreement with the value corresponding to the slowly rotating Kerr metric. The most remarkable result concerns the mass quadrupole moment Q. We found that for a slowly rotating super-compact Schwarzschild star, in the gravastar limit, the relative deviation factor is $\Delta Q/Q\sim 10^{-15}$. These aforementioned results indicate that the external metric of a slowly rotating super-compact Schwarzschild star in the gravastar limit, agrees with the Kerr metric to the requisite order to one part in $10^{15}$. These results provide the long-sought solution to the problem of the source of rotation of the slowly rotating Kerr metric.       
    
\section*{Acknowledgements}
I am grateful with Prof. Pawel O. Mazur for suggesting the problem and invaluable discussions. I would like to thank Profs. Timir Datta, James Knight and Pawel O. Mazur for reading the manuscript and important comments. I am indebted to Dr. Martin Urbanec and specially Prof. John C. Miller for useful suggestions on the numerical part and valuable comments. Finally, I want to thank the Department of Physics and Astronomy at USC where I was supported as a Teaching Assistant during this work.       



\newpage
\bibliographystyle{mnras}
\bibliography{main} 
\bsp	
\label{lastpage}

\end{document}